\documentclass[preprint,showpacs,preprintnumbers,amsmath,amssymb,floatfix]{revtex4}

\usepackage{color}
\usepackage{graphicx}
\usepackage{dcolumn}
\usepackage{bm}
\usepackage{multirow}
\usepackage{multirow}

\definecolor{sienna}{cmyk}{0,0.72,1,0.45}
\definecolor{fg}{cmyk}{0.91,0,0.88,.12}
\definecolor{yellow}{cmyk}{0,0,1,0}
\definecolor{or}{cmyk}{0,1,0.5,0}
\definecolor{magenta}{cmyk}{0,1,0,0}
\definecolor{rubinered}{cmyk}{0,1,0.13,0.45}
\definecolor{blue}{cmyk}{1,1,0,0}
\definecolor{turquoise}{cmyk}{1,1,0,0.5}
\definecolor{aquamarine}{cmyk}{0,1,0,0.0}
\definecolor{midnightblue}{cmyk}{1,0.5,0.0,0.0}
\definecolor{junglegreen}{cmyk}{1,0,0.2,0.5}

\newcommand{\clg}{\color{fg}}

\newcommand{\clb}{\color{blue}}

\newcommand{\clr}{\color{red}}

\newcommand{\clbl}{\color{black}}

\newcommand{\clmb}{\color{midnightblue}}

\begin{document}


\title{The Gradient Mechanism in a Communication Network}
                                                              
\author{Satyam Mukherjee}                                     
\email{mukherjee@physics.iitm.ac.in}                          
\author{Neelima Gupte}                                        
\email{gupte@physics.iitm.ac.in}                              
\affiliation{Physics Department, IIT Madras.}

\date{\today}     
                                                              
\begin{abstract}                                              
We study the efficiency of the gradient mechanism of message transfer in     
a $2-d$ communication network of regular nodes and randomly   
distributed hubs. Each hub on the network is assigned  some
randomly chosen capacity 
and hubs with lower capacities are connected to the hubs with  
maximum capacity. The average travel time of single messages
traveling on this lattice, 
plotted as a function of hub density, shows  q-exponential behavior.
At high hub densities,  this distribution can be fitted well by a 
power law. We also study the relaxation behavior  of  the network when
a large
number of messages are created simultaneously at random    
locations, and 
travel on the 
network towards their designated destinations.
For this situation, in the absence of the gradient mechanism, the network 
can show congestion effects due to the formation of transport traps.
We show that if hubs of high betweenness centrality are connected by the
gradient mechanism, efficient decongestion can be achieved. 
The gradient mechanism is less prone to the formation of traps than
other decongestion schemes. We also study the spatial configurations of transport traps, and 
propose minimal strategies for their elimination. 
                                                              
\end{abstract}


\maketitle
Transport processes on networks have  been a topic of intensive research in recent years. 
Examples of transport processes on networks include the traffic of information packets\cite{tadic,thurn,guilera,rosvall}, transport processes on
biological 
networks \cite{a,c}, and road traffic. The structure and topology of the
network, as well as the mechanism of transport, have been seen to play
crucial roles in the optimization of the efficiency of the transport process\cite{kim}. It is therefore important to study this interplay in the context of realistic networks so their performance can be optimized.    
\newline                                                      
In recent studies it has been shown that transport efficiencies are         
often driven by local gradients of a scalar \cite{danila}. Examples of this     
include electric current and heat flow which are driven by local
gradients of potential and temperature respectively. The gradient
mechanism plays an important role in biological transport processes such    
as cell migration \cite{cell}, chemotaxis, haptotaxis and galvanotaxis. There are
some 
lesser known examples of systems where gradient induced transport on    
complex networks plays an important role. In one example, the computer      
(or a router) \cite{comp} asks its neighbors on the network for their current packet
load. The router balances its load with the neighbor that has the
minimum   
number of packets to route. In this case, the scalar is the negative of     
the number of packets that are routed and a directed flow is induced
along the gradient of the scalar.                             
Recent studies show that the efficiency of transport of gradient
networks   
depends on the topology of the substrate network.             
It has been seen that a gradient                              
network based on a  random graph topology tends to get easily congested,
in the large
network limit. If the substrate network is scale-free \cite{reka}, then the
corresponding gradient network is the least prone to congestion \cite{jamming}.  
A congestion driven gradient, as in the router case, has also been studied \cite{danila}.

In this paper, we study the efficiency of the message transport by the gradient mechanism on two dimensional substrate communication  
network of nodes and hubs. 
A distance based routing procedure for single messages and for many
messages traveling simultaneously in the network are studied in this
context. Single message transport had been considered earlier for
highway traffic\cite{santen}. Both single message and multiple messages
traveling simultaneously in the network have been studied n the context
of communication networks. See e.g. \cite{stout}. We observe that multiple messages traveling simultaneously in the lattice lead to trapping effects. These trapping effects are eliminated by new dynamic strategies.

Communication networks based on two-dimensional
lattices have been
considered earlier in the context of search algorithms \cite{kleinberg} and of network traffic
with routers and hosts \cite{Ohira,Sole,fuks}. The lattice consists of two types
of nodes, the regular or ordinary nodes which are connected 
to each of its nearest neighbors, and the hubs which are connected to all
the nodes in a given area of influence and are randomly disributed in the lattice. 
Thus, the network represents a model with local clustering and geographical separations \cite{warren,cohen}.   
The hubs in the lattice form a random geometry, similar to that of
random geometric graphs, \cite{dall}, 
whereas the ordinary nodes have a regular geometry. 

In the absence of the hubs each node has the same degree of connectivity
and the degree distribution is a ${\delta}$ function with a single peak
at four, the number of nearest neighbors. In the presence of hubs the degree distribution is bimodal \cite{braj}. 
Thus, the degree distribution of this network does not belong to the usual classes, namely the small world \cite{watts} or scale free classes of networks, or to that of random graphs \cite{reka}.
 \newline
The gradient mechanism of message transfer is implemented on this substrate as 
follows. The hubs are randomly distributed on the lattice and the message handling capacities of the hubs are chosen out of a random distribution.
Connections  between hubs are made by the gradient mechanism where the gradient is along the steepest ascent for the capacities associated with the hubs.
The connections between hubs provide short pathways on the lattice, thereby
speeding message transfer. 
In the absence of the gradient mechanism, the average travel time for 
messages traveling between the source and target on the base network plotted 
as a function
of hub density, showed stretched
exponential behavior \cite{braj}. 
If the gradient mechanism is implemented on the lattice, the average
travel time for messages shows $q$-exponential behavior as a function of hub density.
The tail of the distribution shows power law behavior.    
Similar $q$-exponential behavior is observed if the hubs are connected
by random assortative connections, i.e. each hub is connected to two or three
randomly chosen other hubs. The tails of the 
$q$-exponential
 distribution in both cases
show power-law behavior.
This is consistent with the power-law behavior observed earlier at high
hub
densities for  the random assortative connections\cite{braj}.   
The distribution of travel times for the gradient shows 
log-normal behavior. The leading behavior of the travel time
distribution is also log-normal when  the hubs are connected by random
assortative connections, but develops an additive power-law correction. 
We note that log-normal distributions  have been seen for the latency times of the
internet\cite{Sole}, and in
directed traffic flow networks\cite{gautam}.    
In contrast, it is interesting to note that travel-time   
distributions for stationary traffic flow for the  Webgraph   
shows power-law behavior, whereas the Statnet shows $q$-exponential
behavior\cite{tadic1}.

We also study the congestion effects which occur 
when a large number of
messages are run simultaneously on the lattice under distance based
routing. 
Such  effects, which can be seen in    
telephone networks, traffic networks, computer networks and the
Internet \cite{Huberman},
have immense 
practical importance \cite{arenas,moreno,zhao} and are an important
measure of network efficiency. 
Various factors like capacity, band-width and network
topology \cite{Huang}
gives rise to congestion and deterioration of   
the service
quality experienced  by users  due to an increase in network
load. Decongestion strategies, which manipulate factors
like capacity and connectivity to relieve congestion, have attracted
recent attention \cite{braj1,Hui}.  
Since the gradient scheme has proved to be quite efficient at relieving 
congestion in scale free networks \cite{danila}, we test the
efficacy of the gradient scheme for decongestion in our two dimensional   
communication network with the gradient being set up between hubs of
high co-efficients of betweenness centrality \cite{braj1}, and 
compare its success with that other assortative schemes. 
The existence of transport traps has been observed to play a crucial
role in congesting transport on scale-free networks \cite{danila,gallos}. 
Since our network incorporates geographical information, we study the spatial
configuration of traps, the reasons for their formation, and their
contribution to the congestion process. We also propose minimal
strategies for the decongestion of traps. 

The $2-d$ substrate model, and the gradient mechanism for message
transfer  is discussed in section I. We also discuss travel time
distributions and their finite size scaling when a single message travels on the network at a time, in this section.
In section II we study the congestion problem, which occurs 
 when many messages travel on the network
simultaneously. We study the efficiency  of 
a CBC driven gradient with gradient connections between hubs of high
CBC, for decongesting the network. In section III, we study the spatial 
distribution of trapping configurations, their contribution to the
decongestion process and subsequent elimination of traps by different strategies. In each section, we compare the behavior of the gradient mechanism
with that of  other assortative mechanisms. We conclude in the final section.

\section{The gradient mechanism of message transfer}

\begin {figure*}
\begin{tabular}{cc}
\includegraphics[height=6.5cm,width=6.5cm]{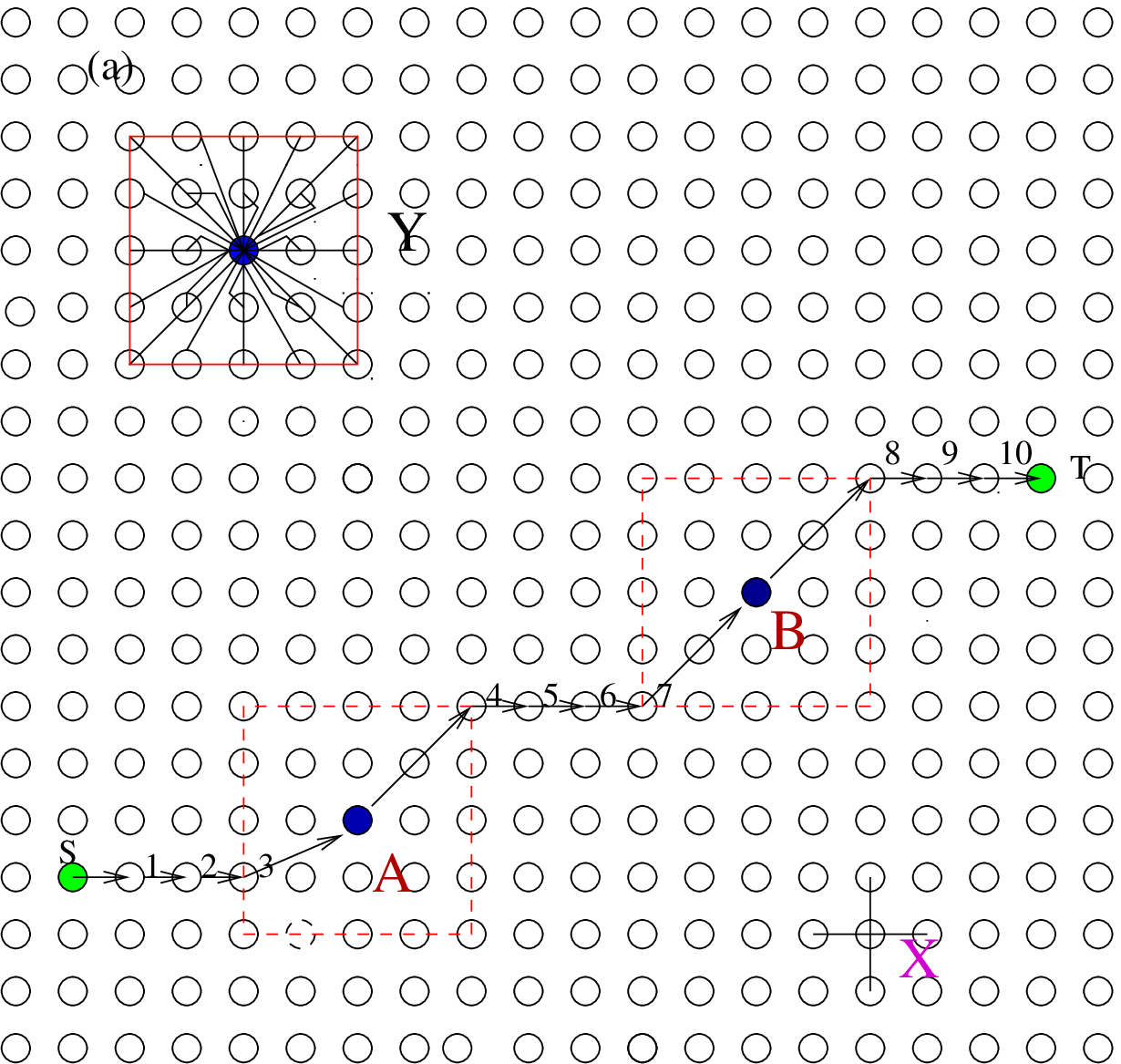}&
\hspace{0.5cm}
\includegraphics[height=6.5cm,width=6.5cm]{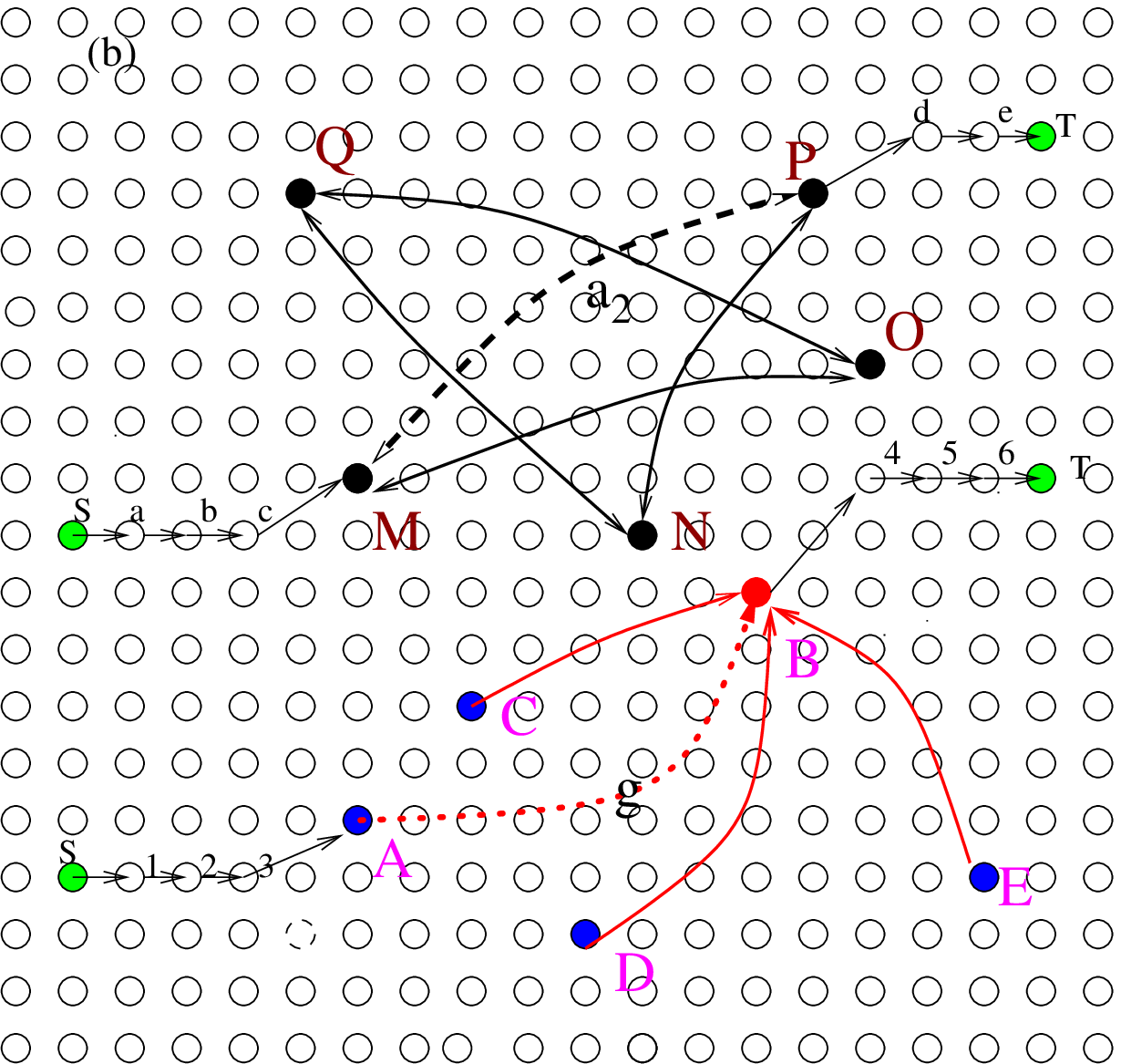}\\
\end{tabular}
\caption{\label{fig:model}(Color online) (a) A two dimensional lattice of $20{\times}20$
nodes. {\it X} is an ordinary node with nearest neighbor connections. Each hub has a square influence region (as shown for the hub {\it Y}). A typical path from the source $S$ to the target $T$ is shown with labeled sites. The path S-1-2-3-A-4-5-6-7-B-8-9-10-T passes through the hubs {\it A} and {\it B}. (b) Hubs $A - E$ are distributed randomly in the lattice and and each hub is assigned with some message capacity between 1 and 10. In the figure $B$ (red circle) has maximum capacity 10. The hubs are connected by the gradient mechanism as shown by one way arrows. After the implementation of the gradient mechanism the distance between {\it A} and{\it B} is covered in one step. The gradient path is given by S-1-2-3-g-8-9-10-T. Hubs $M - Q$ are connected by two way assortative linkages with two connections per hub. A typical path from S to T after the implementation of the two way assortative mechanism between the hubs is shown by S-a-b-c-M-${\bf a_{2}}$-P-d-e-T.}
\end{figure*}

The substrate model on which message communication takes place is shown
in Fig.~\ref{fig:model}(a).
This is  a regular 2-dimensional lattice  
with two types of nodes, the 
regular nodes, connected to their nearest
neighbors (e.g. node $X$ in Fig.~\ref{fig:model}(a)),
and hubs at randomly selected locations which are connected 
to all
 nodes in their area of influence, a square of
side $a$ (e.g. node $Y$ in the same figure). We set free boundary
conditions.  
If a message is routed from a source $S$ to a target $T$
on this lattice through the baseline mechanism,
it takes the path
S-1-2-3-$A$-4-5-6-7-$B$-8-9-10-T as in  Fig.~\ref{fig:model}(a).

To set up the gradient mechanism, we need to assign a  capacity to each
hub, the 
hub capacity being defined to be the number of
messages the hub can process simultaneously. 
Here, each hub is randomly assigned some message capacity
between one and $C_{max}$. 
A gradient flow is assigned from each hub to all the hubs with the maximum 
capacity ($C_{max}$).  
Thus, the hubs with lower
capacities are connected to the hubs with highest capacity $C_{max}$ by the gradient mechanism. In Fig.~\ref{fig:model}(b) if hub {\it A} has capacity
5 and hub {\it B} has capacity 10, then a flow can occur from {\it A} to
{\it B} as shown by the dotted line $g$. Thus the hubs with the highest
capacity $C_{max}$ are maximally connected by the gradient mechanism.
After the implementation of the gradient mechanism, the distance between $A$
and $B$ is covered in one step as shown by the link $g$ and a  message
is routed along the path $S-1-2-3-$A$-$g$-$B$-4-5-6-T$ as shown in Fig.~\ref{fig:model}(b).
Note that gradient mechanism is essentially a one way mechanism (as
shown by $g$).
The same figure, Fig.~\ref{fig:model}(b), also shows the assortative
mechanism  considered earlier for transport on this network\cite {braj}.
Here, each hub is  connected assortatively to two other hubs
randomly chosen from the other hubs. In the assortative scheme a message
is routed along the path S-a-b-c-M-${\bf a_{2}}$-P-d-e-T. The
assortative mechanism, unlike the gradient mechanism, can be one way or two way. We will compare the efficiency of these two schemes 
for single message and multiple messages transport in later sections of this paper.

\subsection{Routing protocol and travel time distributions}

The following routing protocol is followed by messages which travel on
the lattice above. Consider a  message that starts from the source $S$ and travels towards a
target $T$.  
Any node which holds the message at a given time (the current message
holder), transfers the message to the node nearest to it, in the
direction which minimizes the distance between the current message holder
and the target. If a constituent node is the current message holder, it
sends the message directly to its own hub. When the hub becomes the
current message holder, the message is sent to the constituent node
within the square region, the choice of the constituent node being made
by minimizing the distance to the target.  
When a hub in the lattice becomes the current message holder, the
message is transferred to the hub connected to the current message
holder by the gradient mechanism, if the new hub is in the direction of
the target, otherwise it is transferred to the constituent nodes of the
current hub. The constituent node is chosen such that the distance from
target is minimized. If there is degeneracy i.e.,there exists
simultaneously more than one gradient path, we choose the one nearest
to the target. When a message arrives at its target it is removed from
the network. If a message reaches the boundary of the network it remains
at the boundary. 

Two nodes with co-ordinates $(is,js)$ and $(it,jt)$ separated by a fixed
distance $D_{st}$ = $|is-it|$ + $|js-jt|$
are  chosen from a lattice of a given size $L^{2}$, and assigned  to be the
source and target. 
The average travel time for a  message for a fixed source-target
distance,
is a good measure of the efficiency of the network. 
In our simulations,  the  travel time  is calculated for 
a source-target separation of  $D_{st}=142$ on a $100\times100$ lattice,
and averaged over $50$ hub realizations and $1000$ source-target pairs,
with $C_{max}=10$ and $a=3$. These values of $C_{max}$ and $a$ are
retained for all simulations in this paper. 



\begin{figure*}
\begin{tabular}{cc}
\includegraphics[height=6.5cm,width=6.5cm]{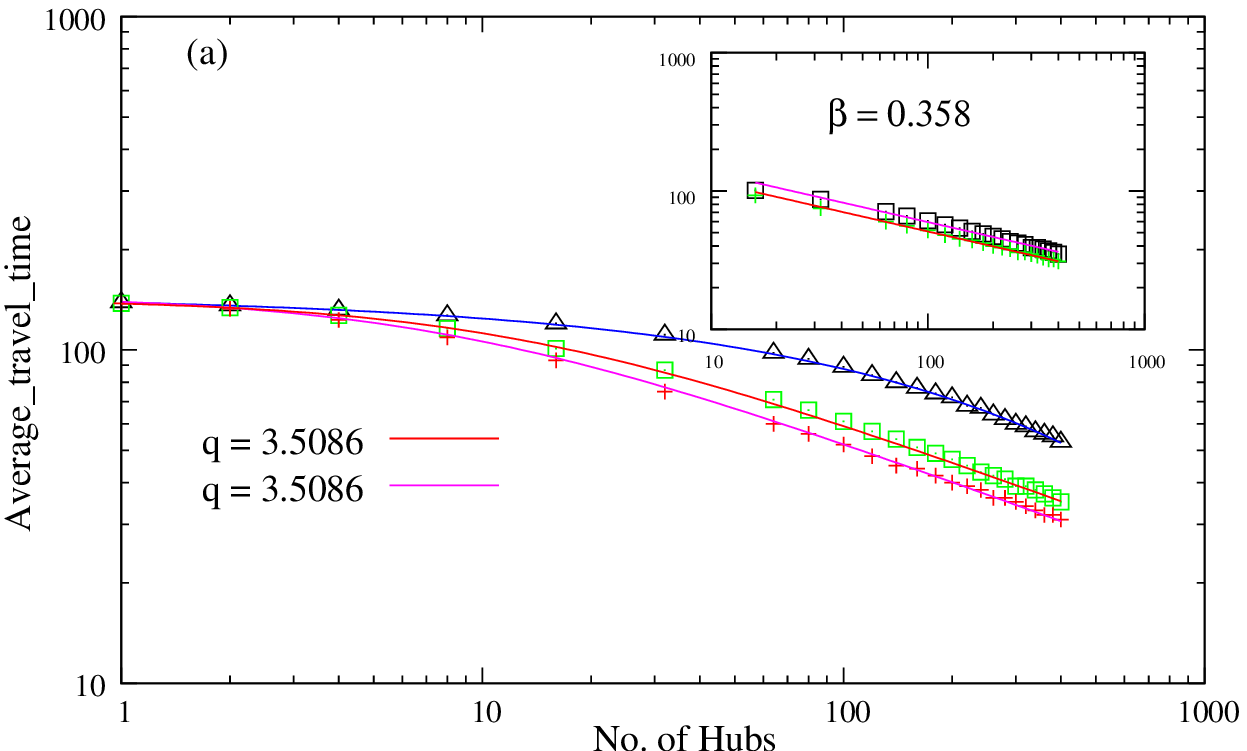}&
\includegraphics[height=6.5cm,width=6.5cm]{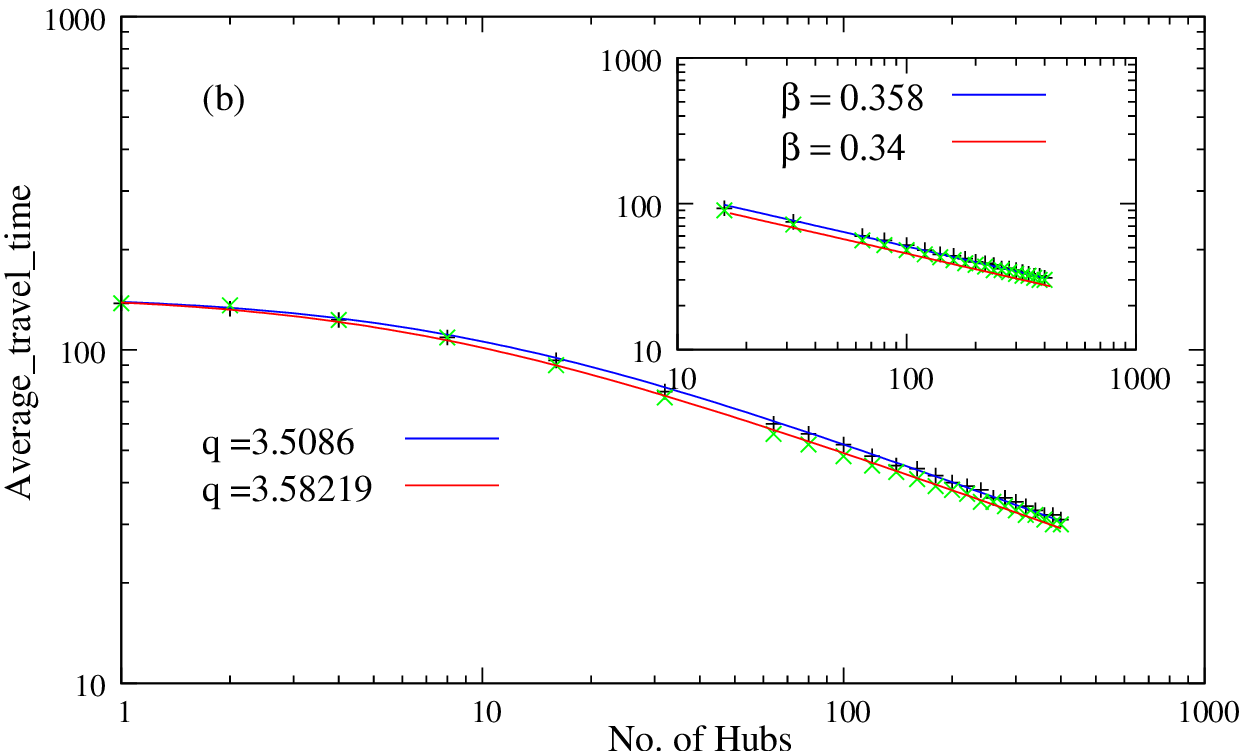}\\
\end{tabular}
\caption{\label{fig:qexp} (Color online) (a) The average travel time as a function of hub density follows a stretched exponential behavior (${\clbl\Delta}$) when the hubs are not connected. If the hubs are connected by gradient as well as one way assortative mechanism, it follows a $q$-exponential behavior. Here $q$ = 3.5086 for the gradient mechanism (${\clg\Box}$) as well as the one way assortative mechanism (${\clr +}$)(b) Here $q$ = 3.5086 for the one way mechanism (${\clbl +}$) and $q$ = 3.58219 for the two way assortative mechanism (${\clg\times}$). As observed, when the hubs are connected by the assortative mechanisms or by the gradient mechanism, the tail of the travel time as function of hub density follows a power law behavior with similar power law exponents. In (a)(inset) for both the gradient mechanism and the one way assortative mechanism $\beta$ = 0.358. In (b)(inset) $\beta$ = 0.34 for the two way assortative mechanism.}
\end{figure*}

The behavior of
average travel time as a function of the number of hubs for a fixed distance
$D_{st}$ between the source-target pairs is plotted in 
Fig.~\ref{fig:qexp} for a 
$100\times100$ lattice and $D_{st}$ of $142$. 
The plot shows data for the original 
network,  as well  for the gradient scheme applied in the network. 
The stretched exponential function $f(x)=Qexp[-Ax^{\alpha}]$, where the 
constants take the values 
${\alpha}$ = 0.50 $\pm$ 0.01132, $A=0.051$ and $Q=146$, gives an
excellent fit to the data. 
However,  the gradient data is fitted best  by the function $f(x)=A(1 - (1 -
q )x/x_{0} )^{(1/(1-q)})$ with  the parameters
 $q =3.5086$, $A=142$ and $x_{0}=0.032$. Thus, the average travel time as a function of the
number of hubs shows $q$-exponential behavior \cite{Tsallis}.

We plot the data for average travel
times for one-way assortative connections in Fig. \ref{fig:qexp}.
The data can be fitted very well by a $q-$exponential function with the
parameters  $q$ = 3.5086. When the number of hubs exceeds 10, the tail of the distribution can be fitted very well by a 
power law $a(x)=P_{a}x^{-\beta}$, where ${\beta}$=0.34 $\pm$ 0.001974
and $P_{a}$=230
(see inset of Fig.~\ref{fig:qexp}(a)), in agreement with the earlier
results. The same inset shows the tail of the travel time distribution
for the gradient case. It is clear that this also fits a power-law 
$g(x)=P_{g}x^{-\beta}$, where ${\beta}$=0.358 $\pm$ 0.006165 and
$P_{g}$=310. Thus, travel times for both the gradient and the one-way assortative
connections show $q$-exponential behavior with tails which can be
approximated by power laws. The values of the $q$-exponents as well
as the values of the exponent $\beta$ of the one way case agree very
well.
We plot the travel time distributions for both one way and
assortative connections in Fig.~\ref{fig:qexp}(b). It is clear that 
both one-way and two-way connections show q-exponential behavior with
power law tails, but the exponents differ slightly as can be seen
from the values in the captions. Thus, the results for the random
assortative connections are in agreement with earlier observations when 
power-law behavior was seen for high hub densities\cite{braj}. 
Earlier studies of networks with growing rules which incorporate 
memory effects have shown $q$-exponential behavior in the degree
distributions \cite{Thurner,Farmer} . It is interesting to note that both the
gradient network and the assortative connections shows a $q$-exponential distribution in the travel times. 
The origin of this behavior may lie in the long range connections
between the hubs.
We also note that the q-exponential does not fit the baseline data for this quantity.

\subsection{Finite Size Scaling}

The dependence of the average travel times as a function of hub density 
discussed above, 
was studied for a
$100\times100$ lattice with a         
$D_{st}$ value of $142$.  However, our results are independent of
lattice size.
We consider this dependence  for lattices of side $L = 300, 500, 1000$ and
corresponding $D_{st}$ values of $424, 712$ and $1420$ respectively. We
consider both the gradient case,
and the case where the connections are assortative.

\begin{figure*}
\begin{tabular}{ccc}
\includegraphics[height=5.8cm,width=5.5cm]{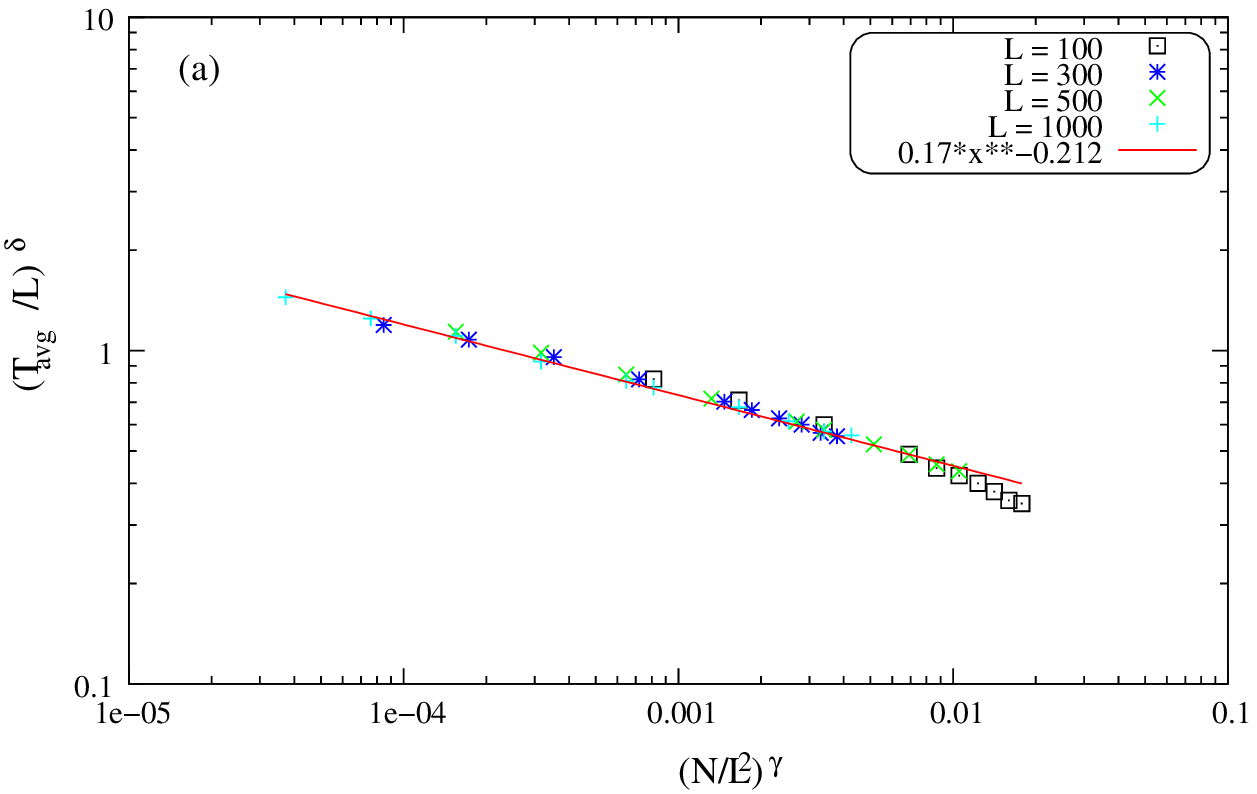}&
\includegraphics[height=5.8cm,width=5.5cm]{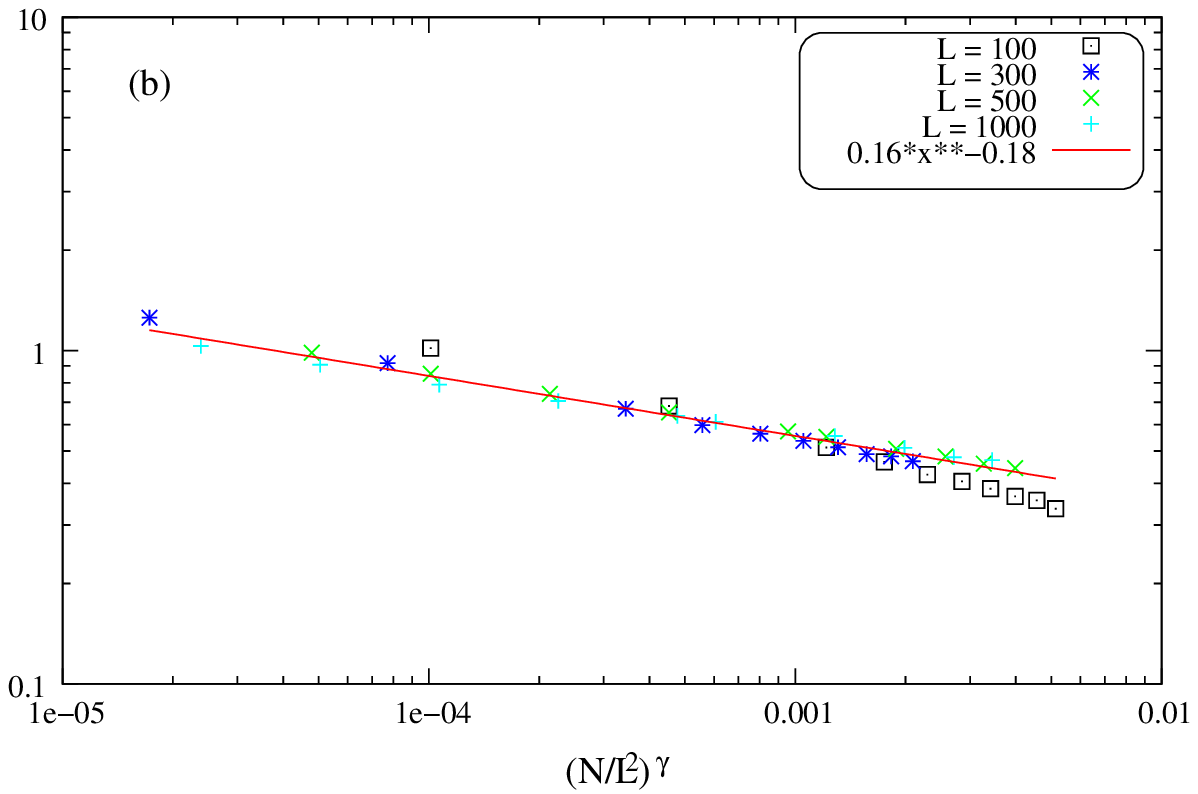}&
\includegraphics[height=5.8cm,width=5.5cm]{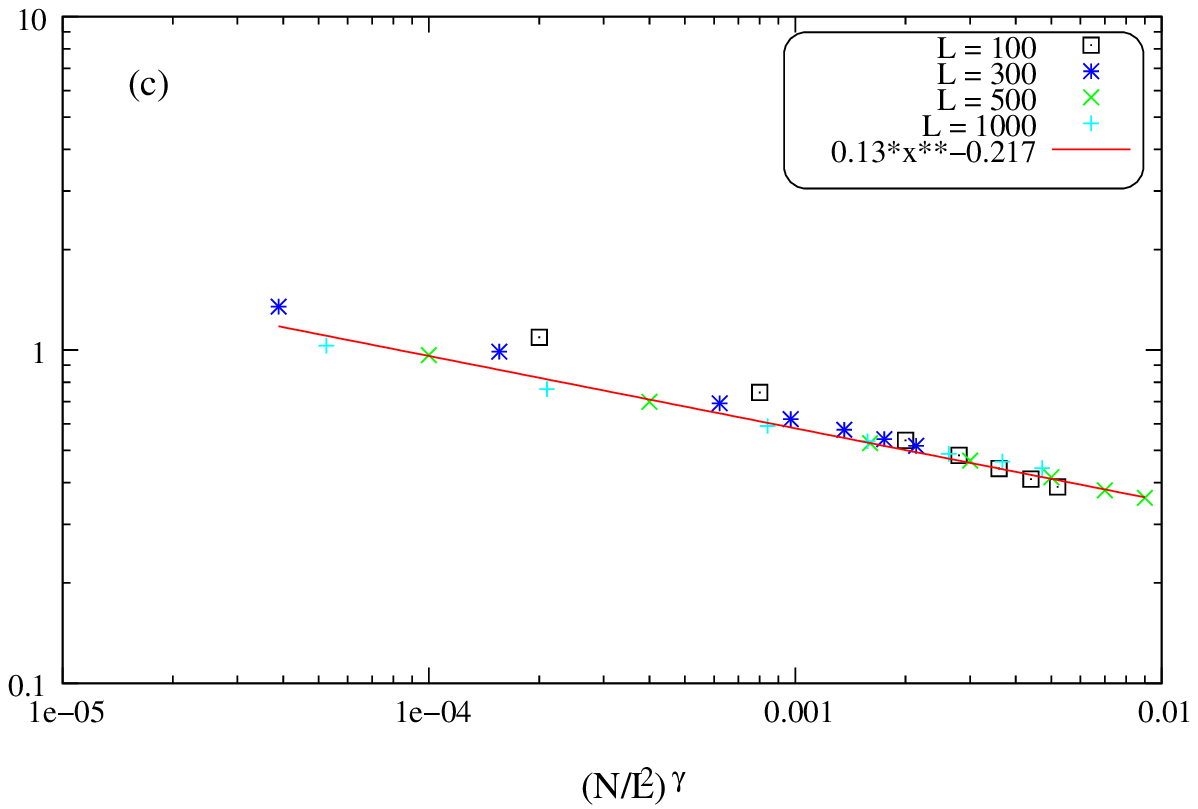}\\
\end{tabular}
\caption{\label{fig:fs} (Color online) The data for L$\times$L two dimensional lattice for different values of L are seen to collapse on top of each other. It is seen that the final curve fits a power law of the form A$x^{-\beta}$. (a) A = 0.17 and ${\beta}$ = 0.212 $\pm$ 0.00224 for the gradient mechanism. (b) A = 0.17  and ${\beta}$ = 0.18 $\pm$ 0.0245 for the two way assortative mechanism. (c) A = 0.13  and ${\beta}$ = 0.217 $\pm$ 0.00275 for the one way assortative mechanism.}
\end{figure*}

Finite size scaling is observed if we plot $(\frac{T_{avg}}{L})^{\delta}$
against
$(\frac{N}{L^{2}})^{\gamma}$
for different lattice size $L$ (Fig. ~\ref{fig:fs}).
Fig.~\ref{fig:fs}(a) plots  
this behavior for the gradient mechanism. It is seen that the data 
observed for different lattice sides $L$ collapse onto each other for
the choice  ${\delta}$ = 1 and ${\gamma}$ = 1.03. The corresponding 
data for the assortative mechanism is shown in Fig.~\ref{fig:fs}(b).
Here the data collapse is observed for ${\delta}$ = 0.88 and ${\gamma}$
= 1.05 for the two way
assortative mechanism, and ${\delta}$=1 and ${\gamma}$ = 1, for the
one way
assortative mechanism.  Thus, the scaling law is,
\begin{equation}
T_{avg} = L^{\mu}f(\frac{N}{L^{2}})^{\frac{\gamma}{\delta}})
\end{equation}
It is
seen that the data for the gradient mechanism scales as a good power law    
up to the scale $(\frac{N}{L^{2}})^{\gamma}$ = 0.001. The power law fit for the     
assortative mechanism is not as good as that for the gradient mechanism     
but scales over a longer stretch with the cut off Fig.~\ref{fig:fs}(b) at
$(\frac{N}{L^{2}})^{\gamma}$ = 0.005.

The distribution of travel times for  messages traveling in the lattice
also shows finite size scaling. 
We considered lattices with  sides $L = 100, 300, 500$ and $1000$ respectively. The
hub density is taken to be $0.5\%$ for all the above cases.

\begin{figure*}
\begin{tabular}{ccc}
\includegraphics[height=5.8cm,width=5.5cm]{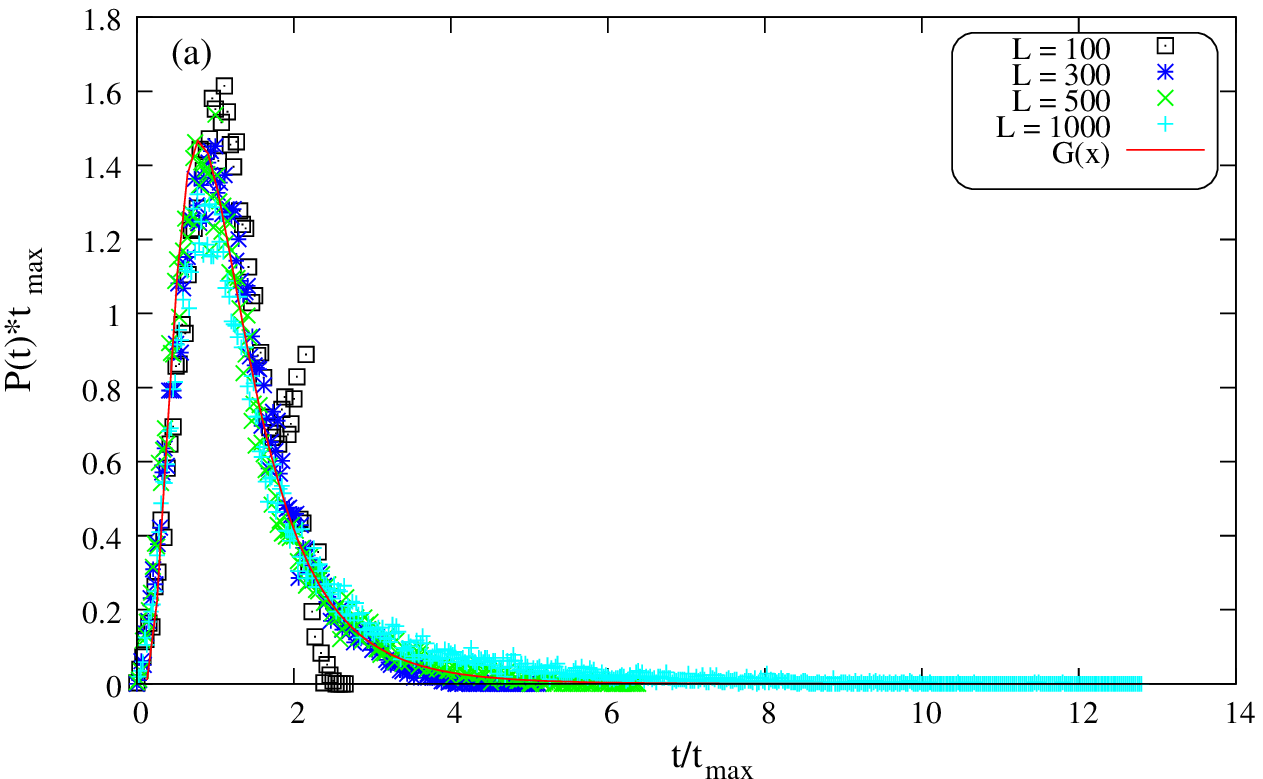}&
\includegraphics[height=5.8cm,width=5.5cm]{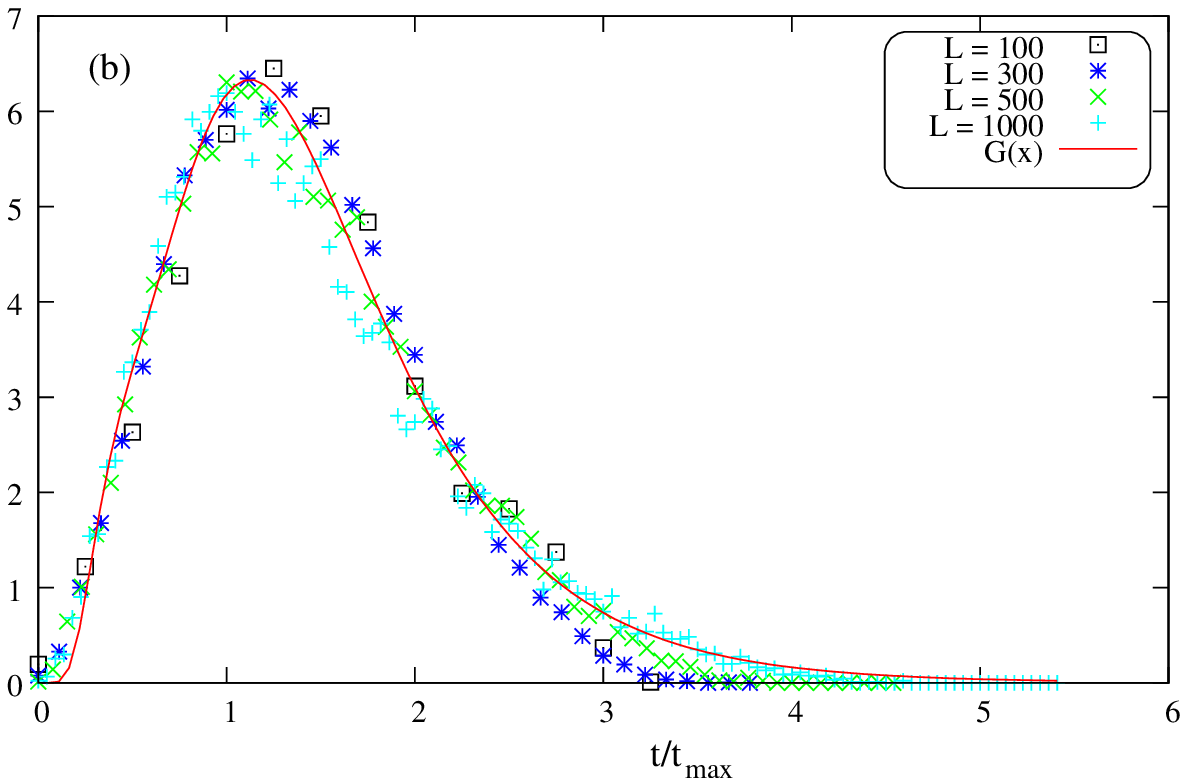}&
\includegraphics[height=5.8cm,width=5.5cm]{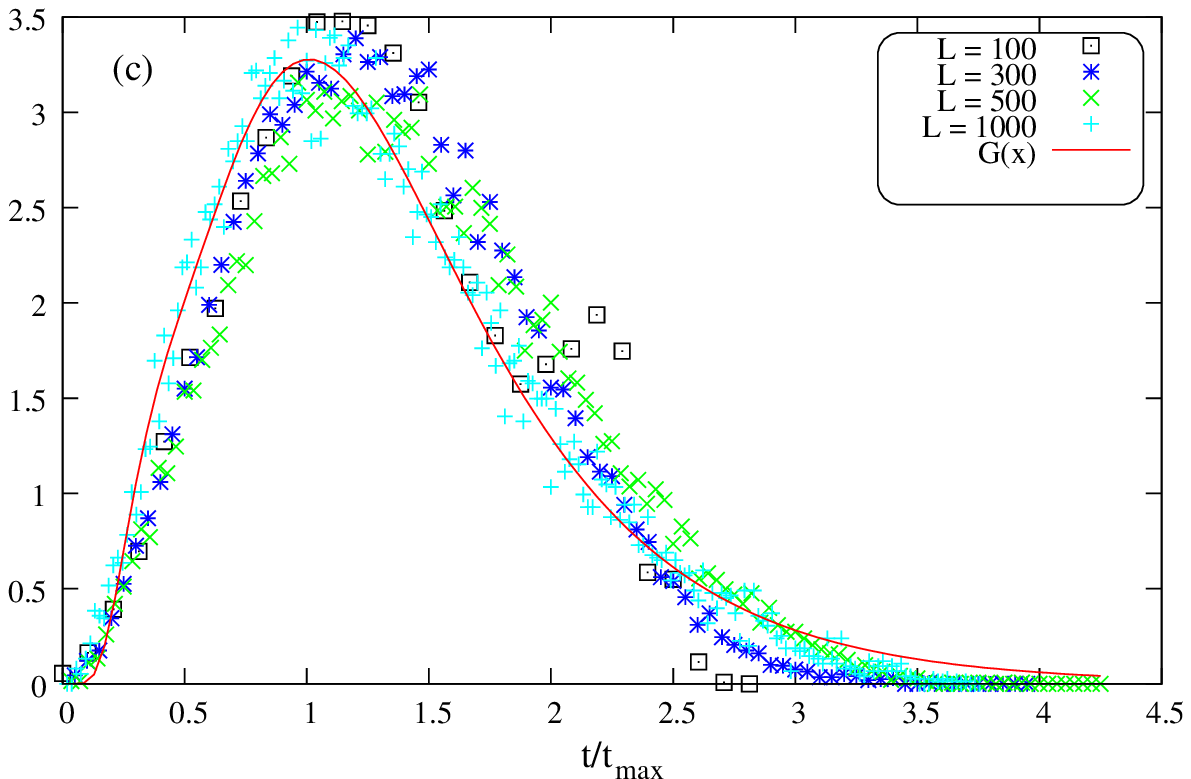}\\
\end{tabular}
\caption{\label{fig:fs2} (Color online)The scaled travel time distribution for (a) the Gradient mechanism, (b) the two way assortative mechanism and (c) the one way assortative mechanism. Different symbols represents lattice of different sizes. (a) The data is fitted by a log-normal distribution (Eq.(3)) where ${\mu}=-1.3$, ${\sigma}=1.336$. The data for (b) and (c) are fitted by a log-normal function with a power law correction (Eq.(4)) (b)${\mu}=-0.080189$, ${\sigma}=1.043$, ${\beta}=4.512$ $\pm$ 0.2012 (c) ${\mu}=-0.3284$, ${\sigma}=1.0767$, ${\beta}=4.25$ $\pm$ 0.1245}
\end{figure*}

The distribution of travel times turns out to have  the  scaling form
\begin{equation}
P(t) = \frac{1}{t_{max}}G(\frac{t}{t_{max}})
\end{equation}
where $t_{max}$ is the value of $t$ at which P(t) is maximum. In a
similar context, it was observed that distribution of optimal path
lengths in random graphs with random weights associated with each link,
has a universal form \cite{Havlin}, but no analytic expression for the universal form was specified. The data
obtained for the gradient (Fig.~\ref{fig:fs2}(a)) can be fitted very
well by a log-normal distribution \cite{gautam} of the form
\begin{equation}
G(x)=\frac{1}{x{\sigma}{\sqrt {2\pi}}}exp(-\frac{({\ln}x-{\mu})^{2}}{2{\sigma}^{2}})
\end{equation}
with ${\mu}=-1.445$, ${\sigma}=1.4744$.
The data obtained for the assortative mechanisms shows longer tails than
the gradient data, and therefore turns out to conform to a   
 a log-normal function with a power law correction of the form
\begin{equation}
G(x)=\frac{1}{x{\sigma}{\sqrt {2\pi}}}exp(-\frac{({\ln}x-{\mu})^{2}}{2{\sigma}^{2}})(1+Bx^{-\beta})
\end{equation}
 where ${\mu}=-0.080189$, ${\sigma}=1.043$, ${\beta}=4.512$ $\pm$ 0.2012 for the two way assortative mechanism (Fig.~\ref{fig:fs2}(b)) and ${\mu}=-0.3284$, ${\sigma}=1.0767$, ${\beta}=4.25$ $\pm$ 0.1245 for the one way assortative mechanism(Fig.~\ref{fig:fs2}(c)). Similar log-normal behavior is obtained for latencies in the internet\cite{Sole} and in the directed traffic flow\cite{gautam}.
We note that the finite size scaling plotted in Fig.~\ref{fig:fs2} is
for a hub density of  
$4 \%$. Similar finite size scaling is observed from hub densities above 
$0.1 \%$. Below this value, we see a bimodal distribution in the travel
times due to the contribution of the nodes and the hubs \cite{braj}, and
no finite size scaling is observed.

\section{Congestion and decongestion}

In the previous section, we studied average travel times, and travel
time distributions, for single messages traveling on the network. In this section we consider a large number of messages which are created at 
the same time and  travel towards their destinations simultaneously.
The hubs on the lattice, and the manner in which they are connected, are the crucial  elements which control the 
subsequent relaxation dynamics, and hence influence the `susceptibility'
of the network.      
On the one hand, it is clear that the hubs provide short paths through the lattice. 
On the other hand, when many messages travel simultaneously on the network,
the finite capacity of the hubs can lead to the  trapping of messages in
their 
neighborhoods,  
and a consequent congestion or jamming of the network. A crucial quantity which identifies these hubs is called
the coefficient of betweenness centrality (CBC)\cite{braj1}, defined 
to
be the ratio of the number of messages $N_k$ which pass through a given
hub $k$ to the total number of messages
which run simultaneously  i.e. $CBC=\frac{N_k}{N}$.           
Hubs with higher CBCs are more prone to congestion. We compare the efficiency of
the gradient mechanism with one-way and two-way assortative CBC mechanisms. We study 
our network in the congested phase where messages are trapped at such
hubs, examine the spatial configurations
of the traps and the success of decongestion strategies.

The gradient mechanism studied here is set up as follows. We choose ${\eta}$ top ranking hubs ranked according
to their CBC values. In the CBC driven gradient mechanism we enhance the
hub
capacities of the ${\eta}$ hubs, proportional to their CBC values by a
factor of ${\kappa}$. The fractional values are set to the nearest
integer values. The hubs are connected by the gradient mechanism. 

We choose $N$  source-target pairs randomly, separated by a fixed distance
$D_{st}$ on the lattice. All sources send messages simultaneously to
their respective targets at an initial time $t=0$. 
The messages are 
transmitted by  
a routing mechanism similar to that for single messages, except
when the next node or hub on the route is occupied.   
We carry out parallel 
updates of nodes.

If the would be recipient node is occupied, then the message
waits for a unit time step at the current message holder.
If the desired node is still occupied after the waiting time is
over, the current node selects any unoccupied node from its remaining
neighbors
and hands over the message.
If all the neighboring nodes are occupied, the message waits at
the current node until one of them is free. If the current message holder is
the constituent node of a hub which is occupied, the message waits at
the constituent node until the hub is free. The rest of the routing
is as described in Section II for single messages. 

In our
simulation we choose ${\eta}$ = 5 and  ${\kappa}$ = 10. We choose a
network of $(100\times100)$ nodes with $N = 2000$ messages and $D_{st}$
=
142. 
It is to be noted that  just $5$ hubs on the lattice have extra connections
in this case, unlike the previous section where every hub has an 
two extra connection. 
The fraction of messages delivered at the end of the run  for given hub
density is shown in Table I. It is clear that the gradient mechanism 
shows a substantial improvement  over the baseline.



\begin{table*}                                                
\caption{\label{tab:table3}The table shows $F$ the fraction of messages
delivered during a run time of $4D_{st}$, as a function of hub density
$D$. The second column shows $F$ for baseline (lattice with unit hub
capacity). The third column shows $F$ for CBC (lattice with augmented
top five hubs). The remaining columns show the fraction of messages
delivered for the gradient mechanism and assortative linkages as
described in the text.}
\begin{ruledtabular}                                          
\begin{tabular}{cccccccc}                                     
D&$F_{Base}$&$F_{CBC}$&$F_{grad}$&$F_{CBC_{a}}$&$F_{CBC_{b}}$&$F_{CBC_{c}}$&$F_{CBC_{d}}$\\
\hline                                                        
0.5 & 0.156 &0.2095& 0.4695 & 0.453 & 0.458 & 0.637 & 0.647  \\    
1.0 & 0.286 &0.405& 0.6185 & 0.588 & 0.567 & 0.7195 & 0.8285  \\  
2.0 & 0.3755 & 0.553 & 0.756& 0.717 & 0.749 & 0.858 & 0.9205  \\   
3.0 & 0.723 & 0.752& 0.9685 & 0.9380 & 0.9395 & 0.9425 & 0.9685  \\
4.0 & 0.903 & 0.868 & 1.0 & 1.0 & 1.0 & 1.0 & 1.0 \\               
\end{tabular}                                                 
\end{ruledtabular}                                            
\end{table*}

Other decongestion mechanism which involve hubs of high CBC have been 
proposed earlier.
It had been observed that introducing assortative
connections between hubs of high $CBC$ has the effect of relieving
congestion \cite{braj1}. This is achieved in two ways : $i)$ One way $(CBC_{a})$ and two way connections $(CBC_{c})$ between the top five hubs ranked by CBC $ii)$ One way $(CBC_{b})$ and two way assortative connections $(CBC_{d})$ between each top five hub and any other hub randomly chosen in the lattice.
In our simulations, the capacity of the top $5$ hubs is enhanced to $5$, so
that these schemes are variants of the CBC scheme. We note that more
than one connection per hub is possible for each one of the two cases.

\begin{figure*}
\begin{tabular}{cc}
\includegraphics[height=6.0cm,width=6.0cm]{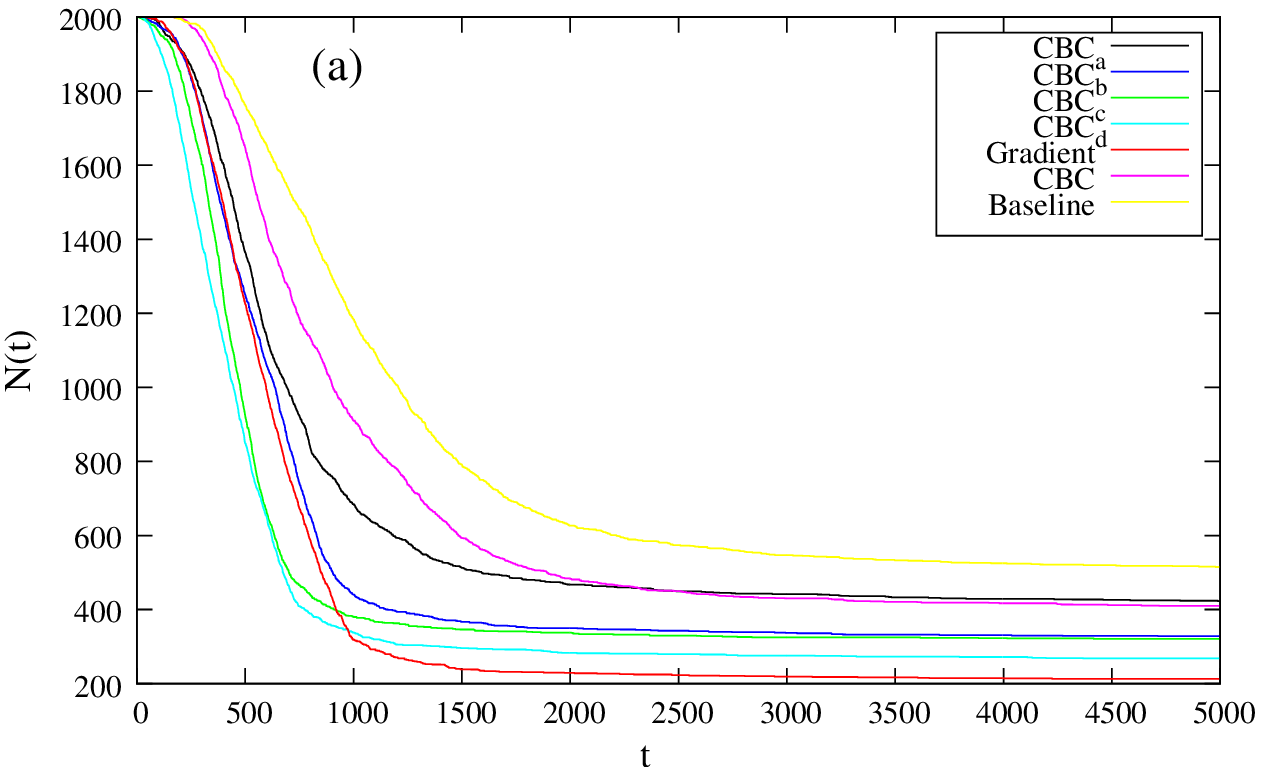}&
\includegraphics[height=6.0cm,width=6.0cm]{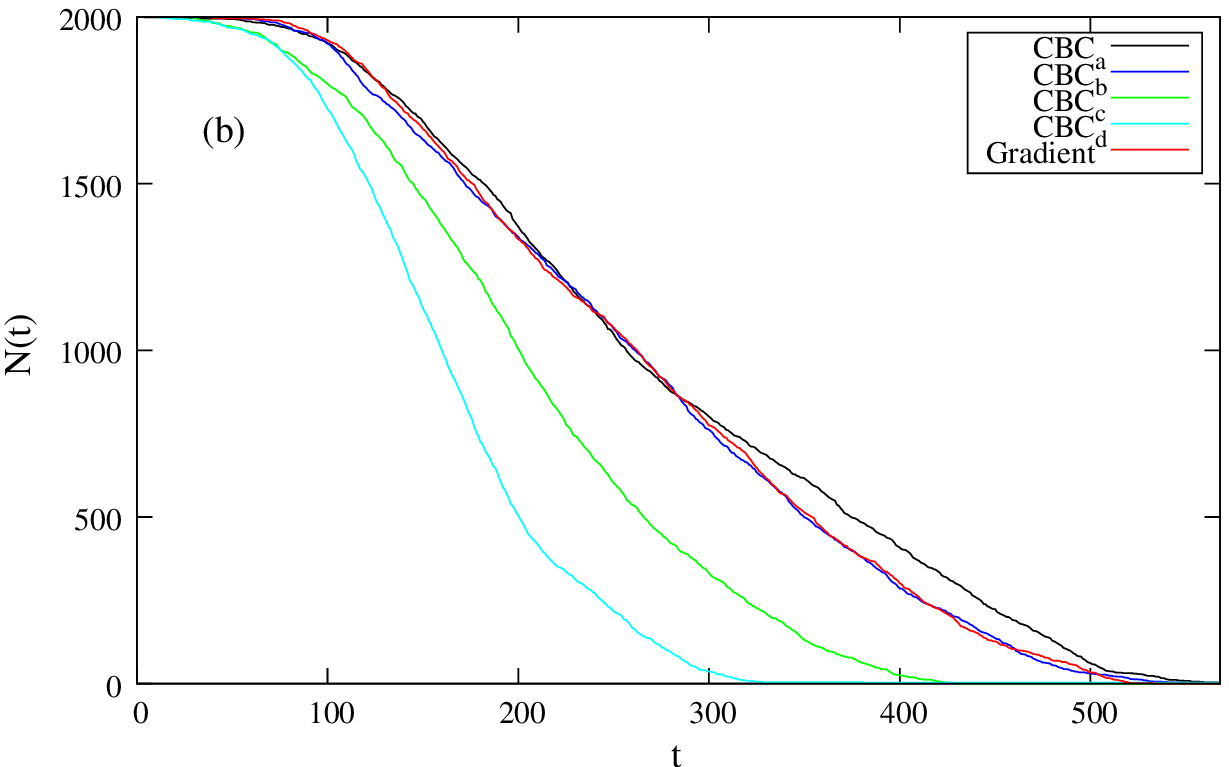}\\
\end{tabular}
\caption{\label{fig:nt} (Color online) Decongestion by different
mechanisms for (a) 50 hubs and (b) 400 hubs in a $100\times100$ lattice
with $D_{st}$ = 142. In (a) run time is set at $5000$ time steps. It is
observed that after a saturation  time $t_s$, the number of messages
undelivered, $N(t)$, gets saturated, indicating the formation of transport traps in the lattice. In (b) all the messages get delivered for the assortative and gradient scheme. The run time is set at $4D_{st}$.}   
\end{figure*}                                                  



It is clear that 
the gradient, which is an inherently one way mechanism, works better
than one way assortative connections of both kinds,
viz.  connections between the 
top five hubs themselves, and the top five hubs and randomly chosen other 
hubs.  
However, it is clear from Table \ref{tab:table3} that the two way connections 
perform better than the gradient at some hub densities. Thus, the gradient is the mechanism of
choice in any set-up where one-way connections are optimal.

The same conclusions can be drawn from 
Fig.~\ref{fig:nt}, which  shows the  plot $N(t)$, 
the number of messages running in the lattice at time $t$,
 as a function
of $t$ for each of these cases, again with 
the parameters  ${\eta}$ = 5 and  ${\kappa}$ = 10, on a 
network of $(100\times100)$ nodes with $N = 2000$ messages, $D_{st}$
=
142 and a run time of 5000 time steps. Fig. ~\ref{fig:nt}(a) is plotted for a hub density of $0.5 \%$, and
Fig. ~\ref{fig:nt}(b) is plotted for a hub density of $4.0 \%$. It is 
clear that all messages get cleared at the higher hub density, whereas 
some messages remain undelivered even after $5000$ time steps at the 
lower hub density. The number of messages remains constant, indicating
that a small fraction of messages have got trapped. It is also
interesting to note that the gradient mechanism is less prone to traps. 
As a result, there is a time at which the gradient mechanism overtakes        
the two way assortative mechanisms in the delivery of messages.
The spatial configuration of traps is interesting. We study this 
in the next section.

\section{Trapping Configurations}

In the decongestion mechanisms discussed above, we studied the transport
of $2000$ messages. For high
hub densities (400 hubs in $100{\times}100$ lattice), all the messages get
cleared when we introduce different schemes of decongestion. Now we
consider 50 hubs in a $100{\times}100$ lattice and a run time of $5000$.
Fig.~\ref{fig:nt}(a) shows that for different decongestion mechanisms, the value of
$N(t)$
saturates after a certain saturation time $t_{s}$. It is observed that one way
assortative mechanisms and the gradient mechanism behave similarly, but the
latter scores over the former in clearing a larger number of messages. A
more detailed analysis of the transport mechanism reveals that the main
cause of non-delivery of messages is the onset of transport traps. 
Similar phenomena of the  formation of traps or congestion nuclei were
observed earlier \cite{alex, danila}. These
traps are formed due to various reasons like the low capacity of high CBC
hubs, the opposing movement of messages from sources and targets situated on
different sides of the lattice,  as well as due to edge effects.
In this section, we look at the spatial configuration of traps under the
various assortative mechanisms.


\begin{figure*}
\begin{tabular}{ccc}
\includegraphics[height=6.5cm,width=5.5cm]{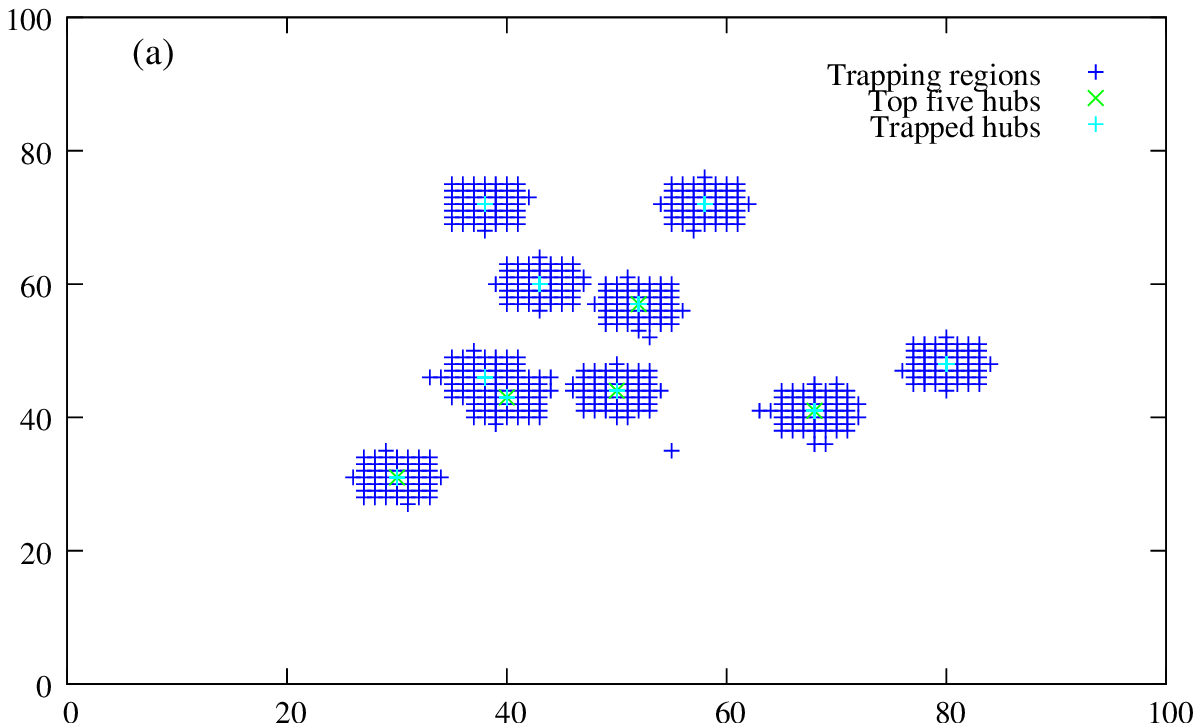}&
\includegraphics[height=6.5cm,width=5.5cm]{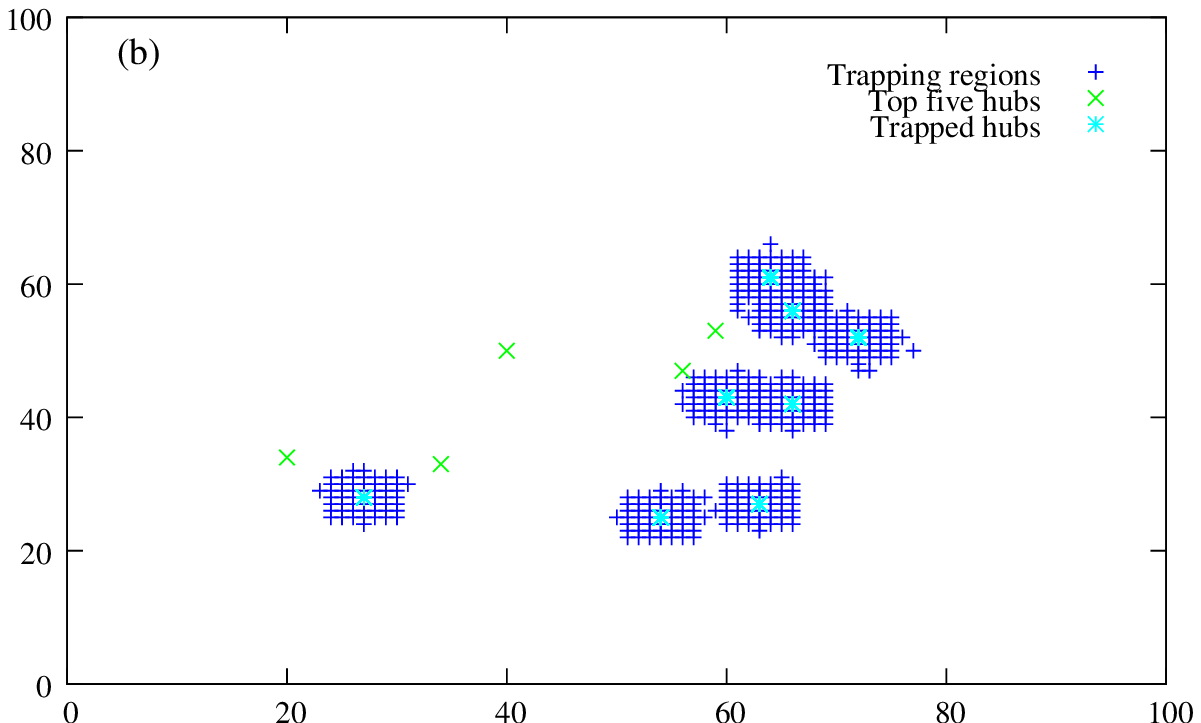}&
\includegraphics[height=6.5cm,width=5.5cm]{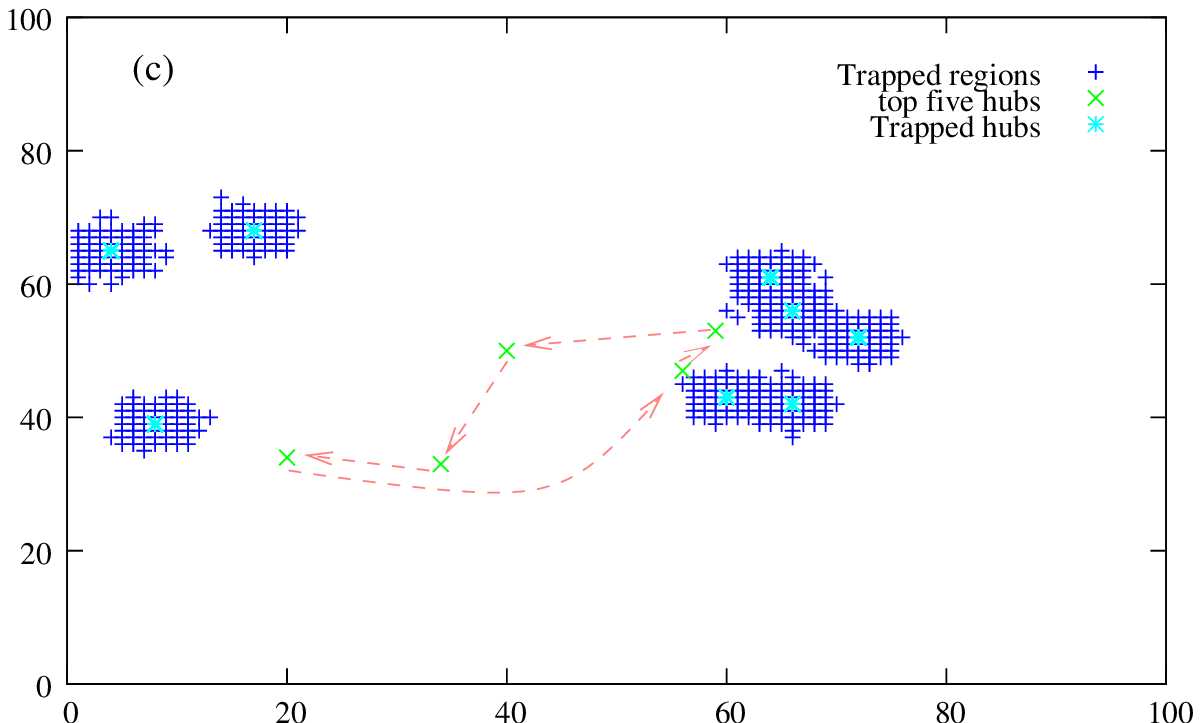}\\
\includegraphics[height=6.5cm,width=5.5cm]{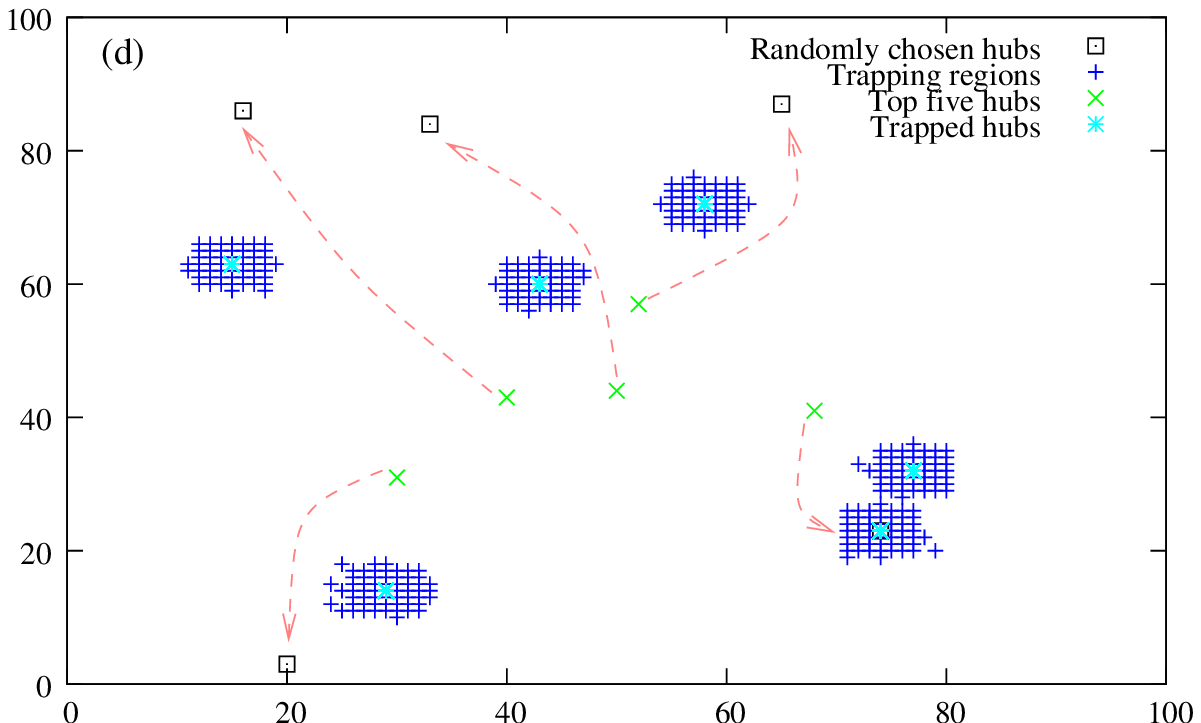}&
\includegraphics[height=6.5cm,width=5.5cm]{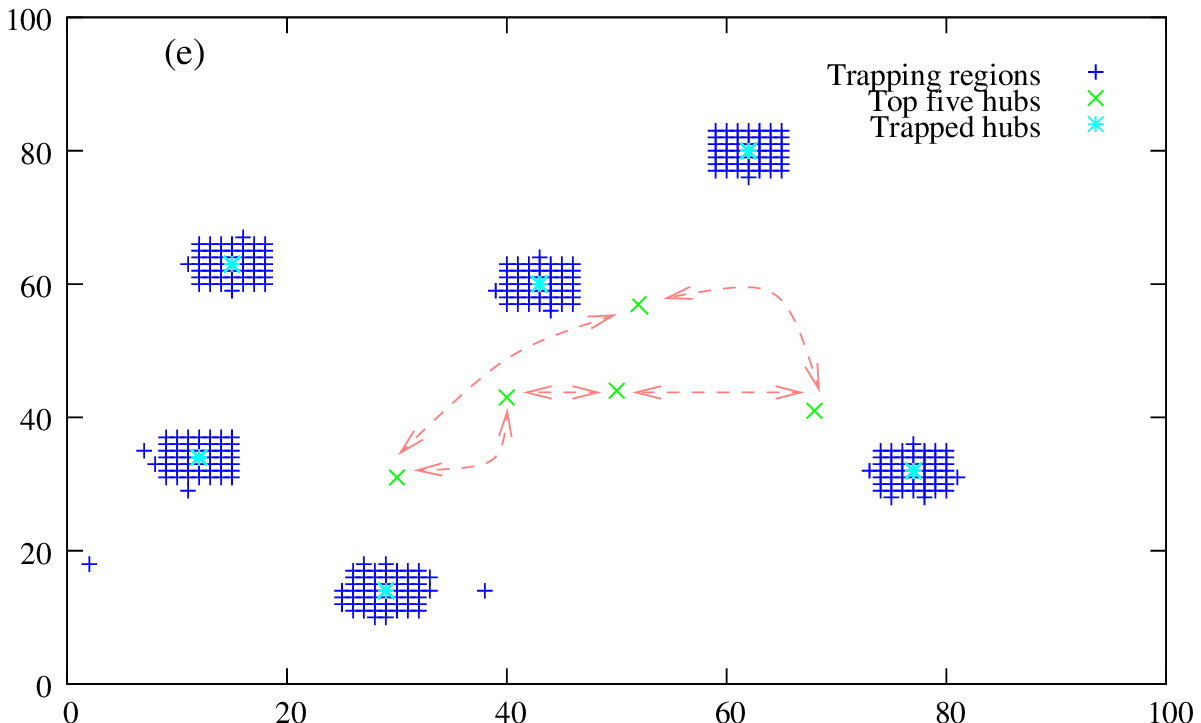}&
\includegraphics[height=6.5cm,width=5.5cm]{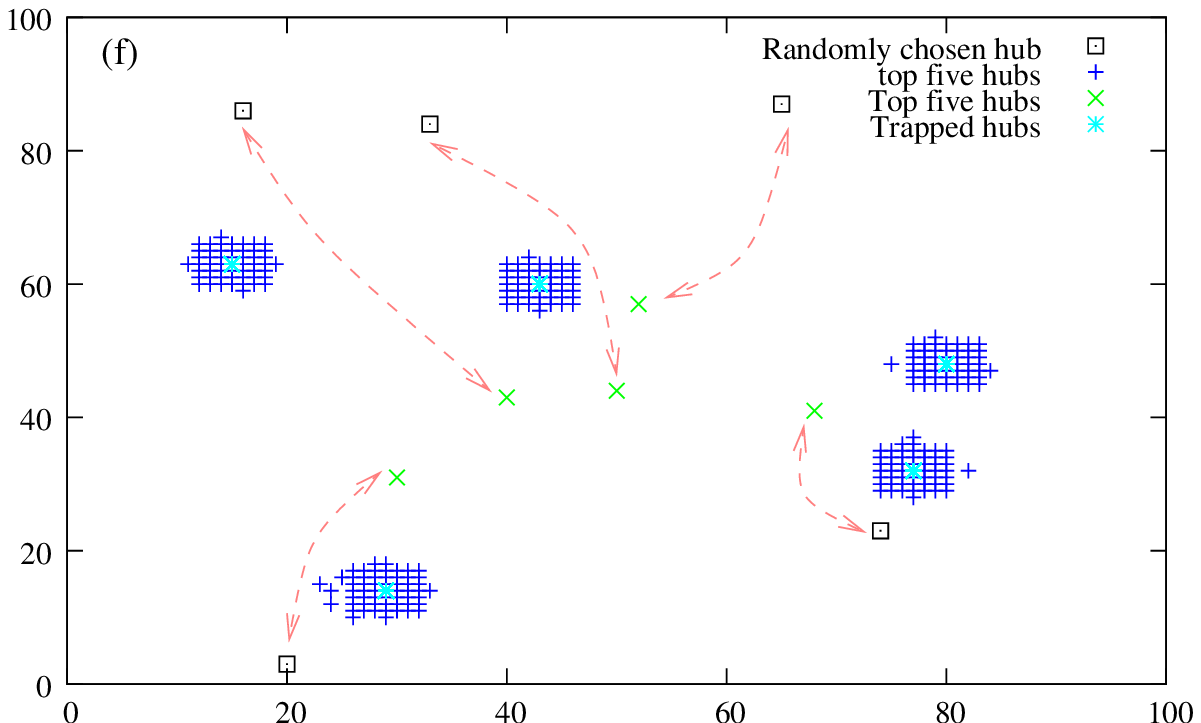}\\
\end{tabular}

\caption{ \label{fig:trap2}(Color online) This figure shows the spatial configuration of traps. The shaded regions ($\clb+$) are the trapping regions where the trapped hubs are indicated by ($\clmb\ast$).The crosses ($\clg\times$) are the top five hubs. The connection between hubs are indicated by one way and two way dashed arrows. The figure shows trapping regions in (a) the baseline mechanism. (b) the CBC mechanism. The top five hubs ($\clg\times$) have enhanced capacity of value 5. (c) the $CBC_{a}$  mechanism. The top five hubs ($\clg\times$) with enhanced capacity and connected by one way assortative mechanism. (d) the $CBC_{b}$ mechanism. Each of top five hubs ($\clg\times$) with enhanced capacity has a one way connection with any other hub chosen randomly ($\Box$) in the lattice. (e) the $CBC_{c}$ mechanism. Top five hubs ($\clg\times$) with enhanced capacity are connected by two way assortative mechanism. (f) the $CBC_{d}$ mechanism. Each of top five hubs ($\clg\times
 $) with enhanced capacity has a two way connection with any other hub chosen randomly ($\Box$) in the lattice.}
\end{figure*}

\begin{table*}
\caption{\label{tab:table1}The table shows the number of hubs trapped and total number of messages trapped in the lattice when different schemes for message transfer are applied. The number of messages trapped in a hub is $(2a+1)^{2}$. Hence total number of messages trapped in the hubs is  $k_{1}(2a+1)^{2}$ where $k_{1}$ is the number of trapped hubs. A Few messages,say $n_{1}$ get trapped in the ordinary nodes adjacent to the constituent nodes of a hub. The total number of messages trapped in the lattice after a given run time is given by $k_{1}(2a+1)^{2}$ + $n_{1}$. We chose 50 hubs in $100\times100$ lattice and a run time of 5000.}
\begin{ruledtabular}
\begin{tabular}{ccccc}
Mechanism&No. of trapped hubs&Messages trapped&Capacity of top five hubs&Saturation time\\
\hline
Baseline & 10 & 515 & 1 & 2500\\
CBC & 8 & 410 & 5 & 2000\\
$CBC_{a}$ & 8 & 413 & 5 & 1500\\
$CBC_{b}$ & 6 & 328 & 5 & 1250\\
$CBC_{c}$ & 6 & 321 & 5 & 1000\\
$CBC_{d}$ & 5 & 268 & 5 & 1000\\
\end{tabular}
\end{ruledtabular}
\end{table*}

\subsection{Transport traps for CBC assortative schemes }

The saturation of messages seen in Fig.~\ref{fig:nt}(a) is a consequence of
the messages getting trapped in high CBC hubs in the lattice. We study
the various geographical trapping patterns formed in the lattice when different decongestion schemes are applied. \newline
In the baseline mechanism due to unit hub capacity all the top five hubs
are trapped (Fig.~\ref{fig:trap2}(a)), and more than $25\%$ of the
messages are trapped (Table \ref{tab:table1}. If the capacities of the top five hubs are
augmented, 
(Fig.~\ref{fig:trap2}(b), CBC in Table \ref{tab:table1}) these hubs get
decongested, but the traps shift to other hubs, and the number of
trapped messages is still large. One way connections between these   
augmented top five hubs 
($CBC_{a}$, Fig.~\ref{fig:trap2}(c))
reduce
the number of trapped messages
quite significantly.  
One way connections between the augmented top five hubs, and randomly
chosen other hubs ($CBC_{b}$, Fig.~\ref{fig:trap2}(e)) do even better. However, 
two way connections, whether among the top five hubs themselves ($CBC_{c}$, 
Fig.~\ref{fig:trap2}(d))
, or among 
the top 5 hubs and randomly chosen other hubs ($CBC_{d}$,
Fig.~\ref{fig:trap2}(f)) work the most efficiently, as can be seen from
the data on the number of trapped hubs and the total number of trapped
messages in Table \ref{tab:table1}.

It is to be noted that no method of reconnecting the hubs
eliminates all traps from the lattice. 
The geographical location of the traps indicates that some of the
trapping is due to edge effects. 
One way connectivity
puts a constraint on the messages which leads to trapping in the
vicinity 
of high CBC hubs. On the introduction of two way connections, this
constraint is removed and the trapping regions can shift to  
other hubs in the lattice.

\begin{figure}
\includegraphics[height=7.5cm,width=7.5cm]{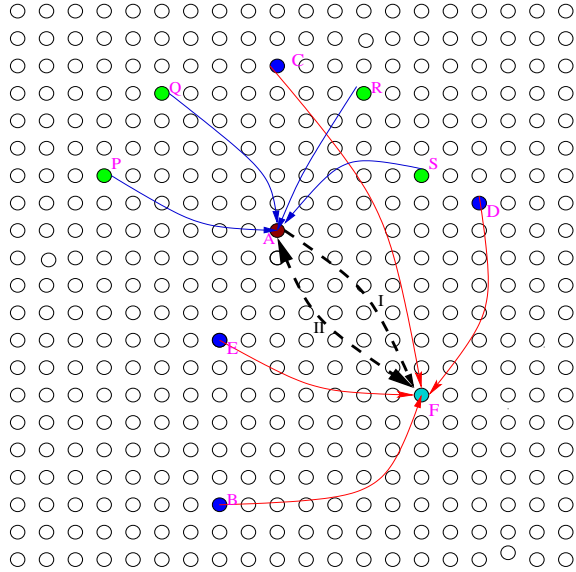}
\caption{ \label{fig:ds1}(Color online) The double star configuration. The hubs B, C, D, E, F are
the top five hubs, and A, P , Q, R, S are the next in order of CBC.
The hubs A and F
are the central hubs for their respective star configurations. The
dotted line I is a one
way connection between A and F. The dotted line II represents a two way connection between A and F. }
\end{figure}

\subsection{Transport traps in  gradient schemes}

We now examine the trapping configurations in different gradient schemes.
We have discussed the single star gradient mechanism in
Fig.~\ref{fig:model}, where the star was formed by applying the 
gradient to the 
top five hubs                                
ranked by CBC
. We
also found that for a long  run time (5000 steps) this mechanism clears
a larger 
number of messages than other decongestion mechanisms, such as the
those discussed above. 
However, a few messages  still remain trapped in this scheme. We try the
double star gradient to  achieve detrapping here.
In the double
star gradient mechanism we form the first star by applying the gradient 
mechanism to the top five hubs
ranked by CBC, and the second star by applying the gradient to  the
next top five hubs.  
We also connect  the double star by applying one-way and two-way
connections between the central hubs as shown in Fig.~\ref{fig:ds1}

\begin{figure*}
\begin{center}
\begin{tabular}{cc}
\includegraphics[height=6.5cm,width=6.5cm]{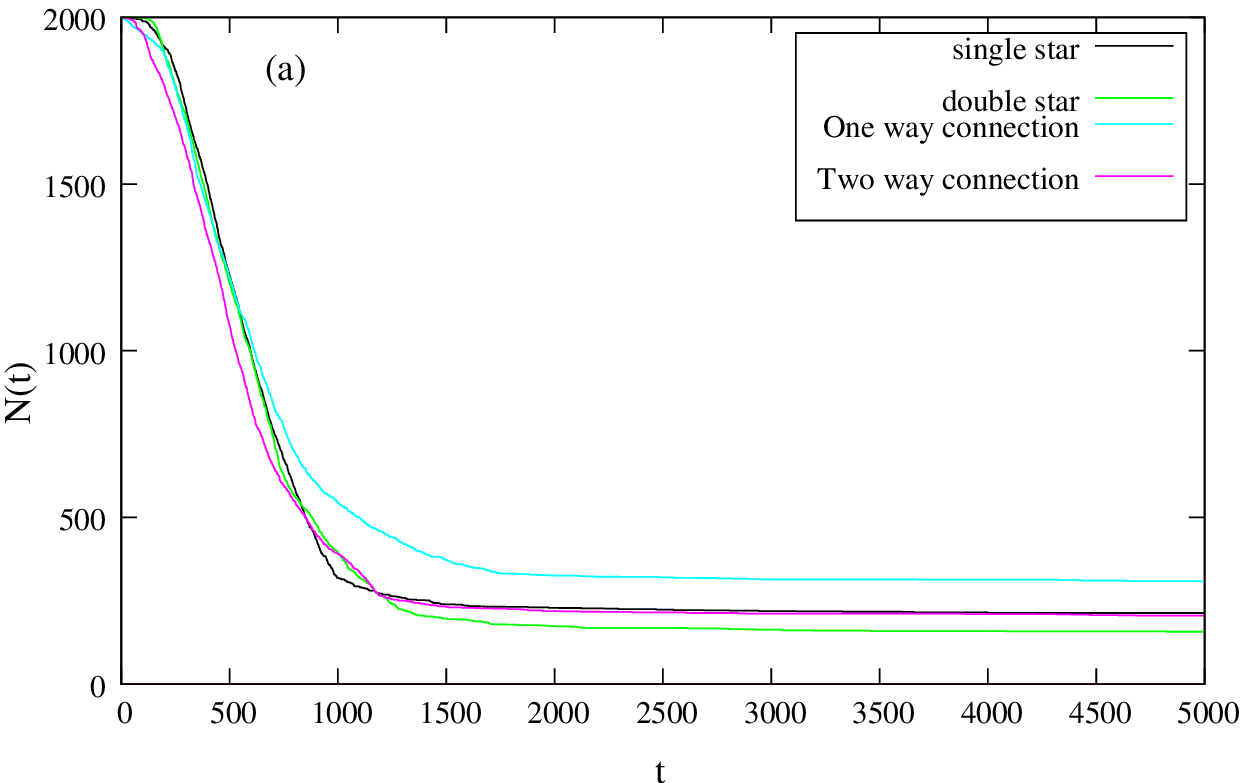}&
\includegraphics[height=6.5cm,width=6.5cm]{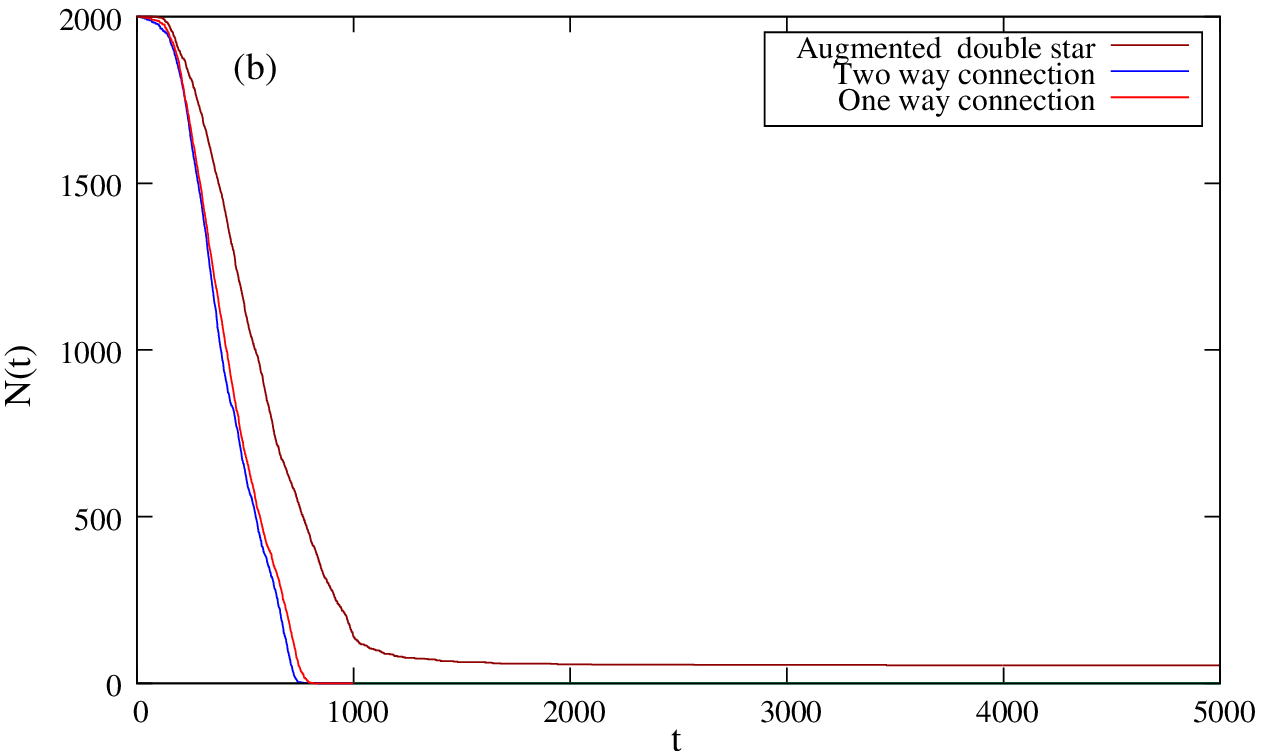}\\
\end{tabular}
\caption{ \label{fig:ds3}(Color online) (a) Trapping in different types of gradient
schemes for 50 hubs in a $100\times100$ lattice.The run time is set at 5000.
(b) Messages clear faster when the capacity of the central hubs of
double star configuration are augmented. Total decongestion occurs when we introduce one way and two way
connections between the two central hubs of the double star and double
their capacities to that of their original values. We considered 50 hubs
in $100\times100$ lattice.}
\end{center}
\end{figure*}

\begin{figure*}
\begin{center}
\begin{tabular}{cc}
\includegraphics[height=6.5cm,width=6.5cm]{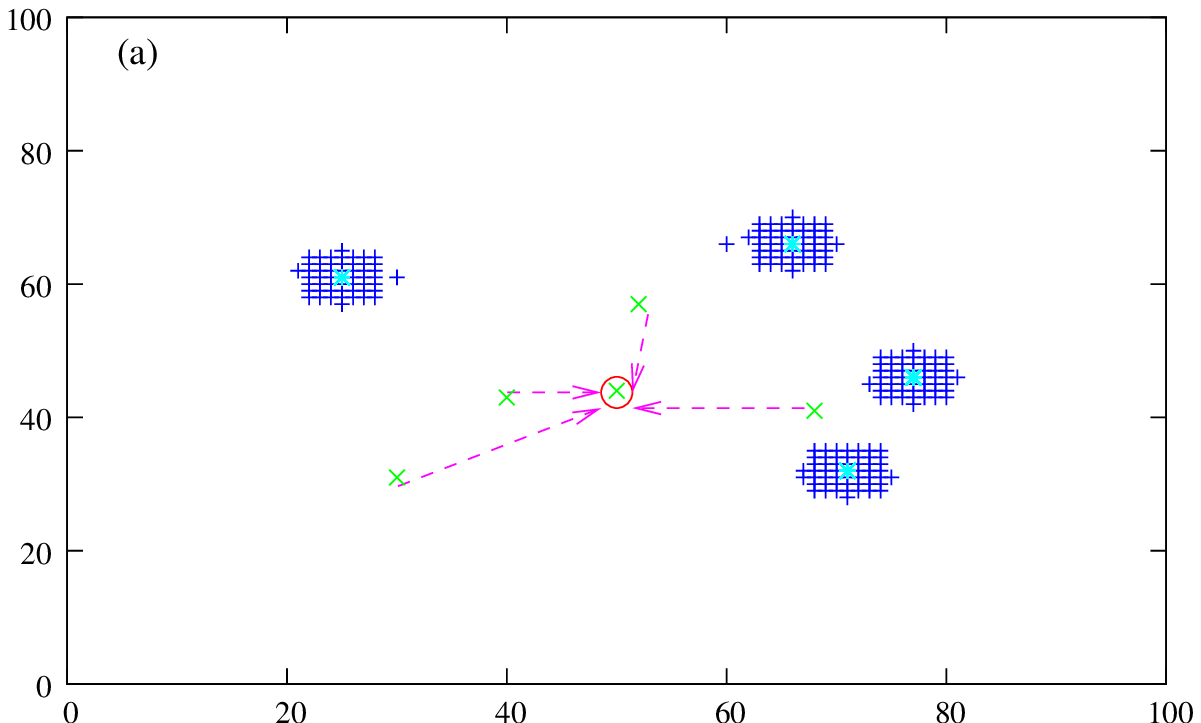} &
\includegraphics[height=6.5cm,width=6.5cm]{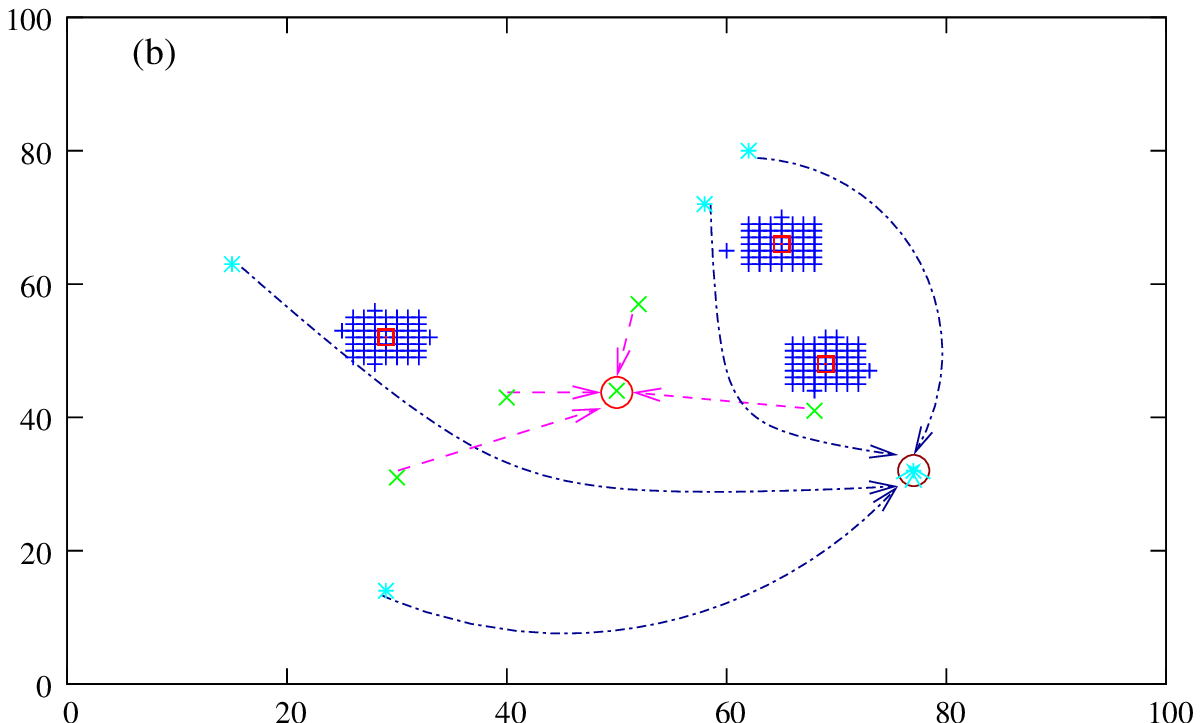}\\
\includegraphics[height=6.5cm,width=6.5cm]{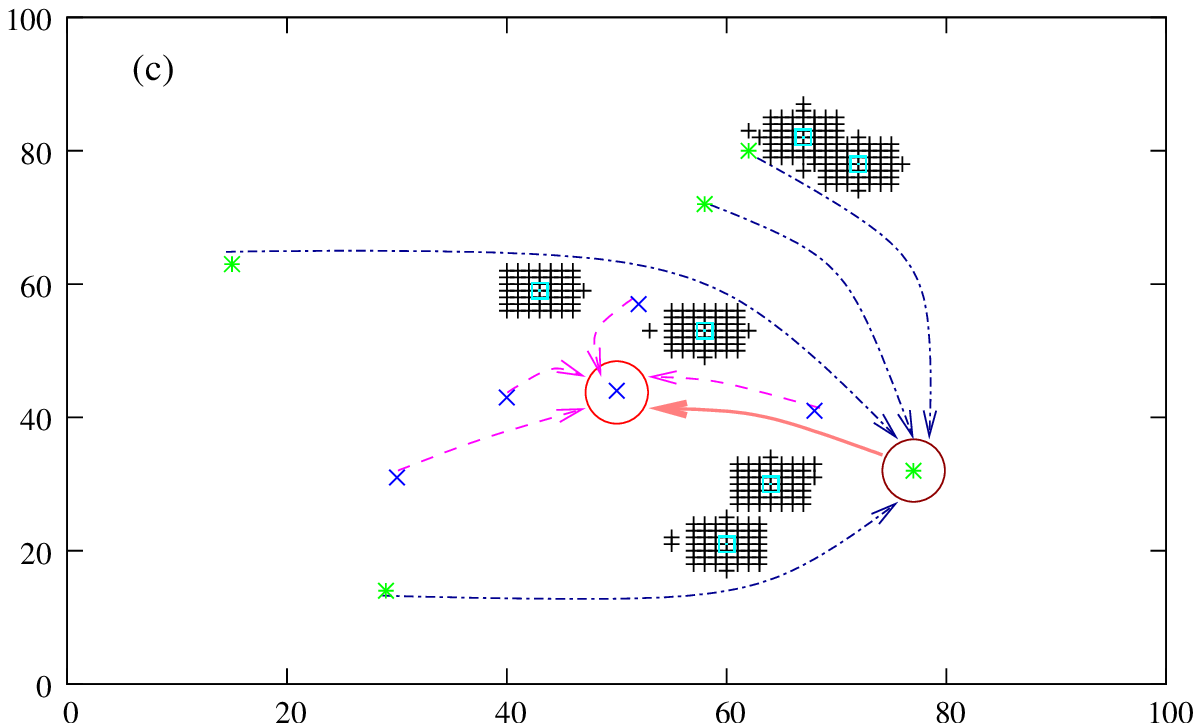}& 
\includegraphics[height=6.5cm,width=6.5cm]{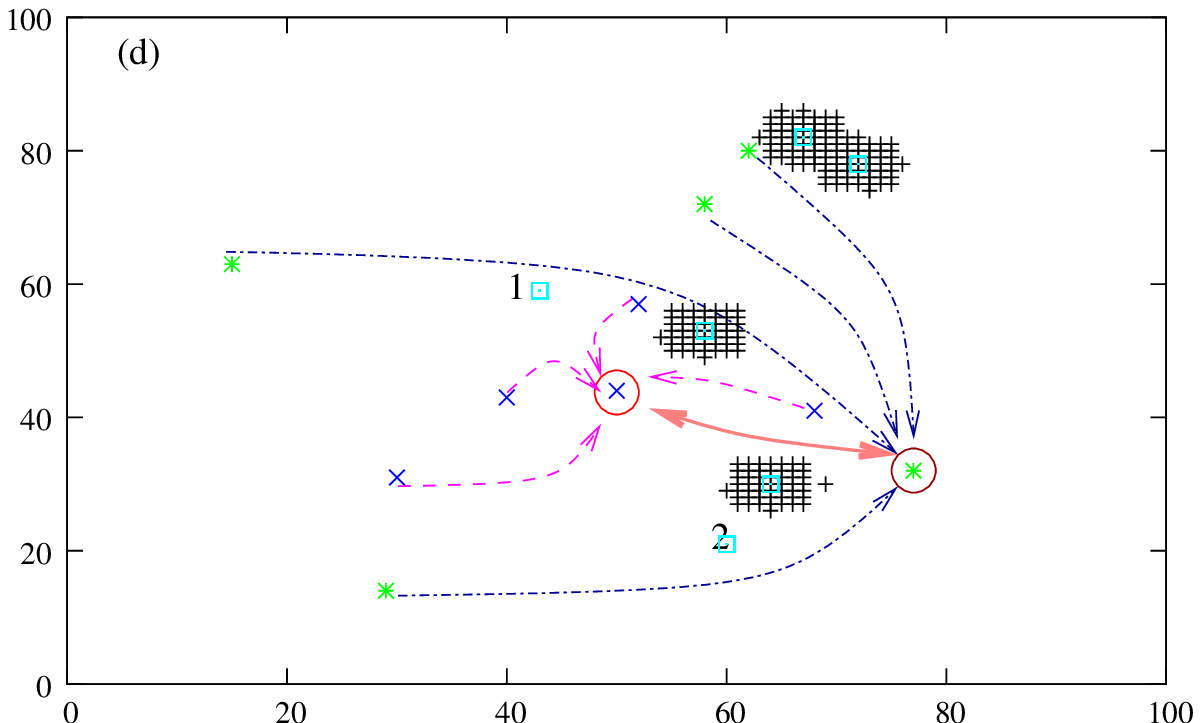}\\
\end{tabular}                                                 
\caption{\label{fig:ds2}(Color online) Trapping regions for 50 hubs in a $100\times100$ lattice in (a) Single star gradient mechanism .The capacities of the top 5 hubs ($\clg\times$) are distributed proportional to their CBC values with a multiplicative factor of 10. The messages are trapped in the hubs ($\clmb\ast$)(b) Double star gradient mechanism. The capacities of the top five hubs ($\clg\times$) and the next top five hubs ($\clmb\ast$ ) are distributed proportional to their CBC values with a multiplicative factor of 10. The patches ($\clb+$) indicate trapped regions which are trapped hubs ($\clr\Box$). The number of patches are less than that of single star configuration. (c) The one way connection between the two central hubs ($\clb\times$ and $\clg\ast$) of double star configuration. The shaded regions (+) are the trapping regions in the lattice,where the trapped hubs are indicated by ($\clmb\Box$). (d)The two way connection between the two central hubs ($\clb\times$ 
 and $\clg\ast$) of double star configuration . Hubs 1 and 2 are cleared when a two way connection between the central hubs are introduced. }
\end{center}                                                  
\end{figure*}

Fig.~\ref{fig:ds2} shows the trapping regions in the double star
configuration. 
The double star configuration (Fig.~\ref{fig:ds2}(b)) clears messages faster than
the single star configuration (Fig.~\ref{fig:ds2}(a)) due to the presence of additional short
cuts. 
If the capacity of the central hubs of the double star is doubled,
some messages still remain undelivered. (See 
Table~\ref{tab:table2}) as well as Fig. \ref{fig:ds3}.
If one way assortative connections (Fig.~\ref{fig:ds2}(c)) are added between the  central 
hub of each star, messages get trapped in the vicinity of the central
hubs. 
The introduction of two way connections (Fig.~\ref{fig:ds2}(d)) improves the situation,
but still does not clear the congestion completely. 
In order to clear the congestion completely, we need to enhance the capacity of the
central hubs of the two stars (Table~\ref{tab:table2}) as well as add an
assortative connection between the two central hubs. If the capacity of these two central hubs
is doubled relative to their original capacity, all the messages get
cleared for both the assortative one way and two way cases
(Fig.~\ref{fig:ds3}(b)). Thus, the capacity of the central hubs of the
stars remains the limiting factor in the clearing of congestion.  
However, due to the optimal nature of the double star with two
connections configuration, increase of capacity  at just the two 
central hubs of the star is sufficient to relieve congestion.

\begin{table*}
\caption{\label{tab:table2}The table shows the number of hubs trapped and total number of messages trapped in the lattice when different CBC driven gradient schemes for message transfer are applied. We chose 50 hubs in $100\times100$ lattice and a run time of 5000. The position of central hub for single star configuration is (50,44) and that for double star configuration is (50,44) and (29,14).}
\begin{ruledtabular}
\begin{tabular}{ccccc}
Gradient &No. of &Messages &Saturation & Capacity of \\
Mechanism & trapped hubs &trapped &time &central hubs \\
\hline
\\
\multirow{1}{*}{Single Star configuration} &\multirow{1}{*}{4}&\multirow{1}{*}{213}&\multirow{1}{*}{1200}

                             		   & 10\\

\multirow{1}{*}{Double Star (D.S) configuration} &\multirow{1}{*}{3}&\multirow{1}{*}{157}&\multirow{1}{*}{1400}
					   & 10,4 \\
\\
\multirow{1}{*}{Double Star (D.S) one way} &\multirow{1}{*}{6}&\multirow{1}{*}{308}&\multirow{1}{*}{1800}
					   & 10,4\\ 
\multirow{1}{*}{Double Star (D.S) two way} &\multirow{1}{*}{4}&\multirow{1}{*}{205}&\multirow{1}{*}{1250}

                             		   & 10,4\\
\\
\multirow{1}{*}{Augmented D.S.} &\multirow{1}{*}{1}&\multirow{1}{*}{54}&\multirow{1}{*}{1200}
					   & 20,8\\
\multirow{1}{*}{Augmented D.S. 1 way} &\multirow{1}{*}{0}&\multirow{1}{*}{0}&\multirow{1}{*}{-}
					   & 20,8\\ 
\multirow{1}{*}{Augmented D.S. 2 way} &\multirow{1}{*}{0}&\multirow{1}{*}{0}&\multirow{1}{*}{-}
					   & 20,8\\
\\
\end{tabular}
\end{ruledtabular}
\end{table*}

\subsection{Elimination  of trapping effects}

As seen above, the occurance of transport traps is unavoidable in
networks which incorporate hubs. On the other hand,
the existence of hubs is essential for providing short paths and short
travel times on the network. In the case of the gradient mechanism, 
the elimination of trapping effects in the double star configuration needed a
combination of addition of connectivity, as well as capacity enhancement. 
This is a static strategy. Static and dynamic strategies of message routing have been considered earlier for communication networks\cite{raghu}. In this section we outline two dynamic
strategies of eliminating trapping effects. One involves capacity enhancement,
and the other involves rerouting. The new strategies are invoked after
the number of messages which reach the target has saturated, that is 
at times which exceed $t_{s}$.    

\begin{enumerate}

\item In Strategy-I, we enhance the capacity of the temporary targets of
the trapped messages by unity. The number of messages running on the
lattice at time $t$ as a function of time can be found in  Table
\ref{tab:table4}. Each column is labeled by the nature of the substrate
network on which the messages run. The traps clear very fast (within 200
time steps) 
as can be seen from the table, despite the enhancement of capacity being
small. The baseline clears the slowest, and the gradient the fastest.

\item In Strategy-II, we bypass the transport  traps  by sending the
messages which will encounter traps by a different route. 
If the temporary target of a given message turns out to be a transport trap, 
the message is assigned a different temporary target. The newly
assigned temporary target is chosen along the direction of the final
target (Table \ref{tab:table5}). This strategy is not efficient for the
baseline mechanism. For this case, messages start clearing only to be trapped again
after a certain time. However, this strategy acts very fast when applied
to  the  assortative network  and the gradient
network. 

\end{enumerate}

Its observed that in terms of rate of delivery of messages, Strategy-$I$
is more efficient than Strategy-$II$ when applied on the baseline, the
$CBC$, the one way assortative mechanisms and the gradient mechanism.
Strategy-$II$ performs better than Strategy-$I$ when applied on the two
way assortative mechanisms, since the number of alternate paths is
larger.    

\begin{table*}                                                
\caption{\label{tab:table4} The table shows the comparison of values of
$N(t)$ at time $t$, for various decongestion schemes when Strategy-$I$ is applied. }
\begin{ruledtabular}
\begin{tabular}{cccccccc}
$t$&$N_{Base}$&$N_{CBC}$&$N_{CBC_{a}}$&$N_{CBC_{b}}$&$N_{CBC_{c}}$&$N_{CBC_{d}}$&$N_{Grad}$\\
\hline
4800& 517 & 410 & 425 & 328 & 321 & 268 & 213\\
4850& 474 & 329 & 274 & 226 & 182 & 151 & 155\\
4900& 184 & 59 & 57 & 71 & 42 & 20 & 27 \\
4950& 20 & 1 & 3 & 2 & 3 & 2 & 0 \\
5000& 1 & 0 & 0 & 0 & 0 & 0 & 0 \\

\end{tabular}
\end{ruledtabular}
\end{table*}

\begin{table*}                                                
\caption{\label{tab:table5} The table shows the comparison of values of
$N(t)$ at time $t$, for various decongestion schemes when Strategy-$II$ is applied.}
\begin{ruledtabular}
\begin{tabular}{cccccccc}
$t$&$N_{Base}$&$N_{CBC}$&$N_{CBC_{a}}$&$N_{CBC_{b}}$&$N_{CBC_{c}}$&$N_{CBC_{d}}$&$N_{Grad}$\\
\hline
4800& 517 & 410 & 425 & 328 & 321 & 268 & 213\\
4850& 513 & 397 & 398 & 308 & 155 & 114 & 172\\
4900& 450 & 301 & 285 & 184 & 24 & 11 & 50 \\
4950& 323 & 200 & 189 & 47 & 4 & 1 & 1 \\
5000& 247 & 98 & 58 & 4 & 2 & 0 & 0 \\
5050& 209 & 9 & 1 & 0 & 0 & 0 & 0 \\
5100& 187 & 0 & 0 & 0 & 0 & 0 & 0 \\

\end{tabular}
\end{ruledtabular}
\end{table*}

\section{Conclusion}

To summarize, in this paper, we have studied network traffic dynamics for single message and multiple message transport in a communication network of nodes and hubs which incorporates geographic clustering. The gradient is implemented by assigning 
each hub on the network   some
randomly chosen capacity and connecting 
hubs with lower capacities  to the hubs with
maximum capacity.

The average travel time of single messages
traveling on this lattice,
plotted as a function of hub density, shows  stretched exponential
behavior for the base network in the absence of  the gradient,
but shows  q-exponential behavior
with a power-law tail at higher hub densities, if the hubs are connected
by the gradient mechanism. 
A similar q-exponential distribution with a power-law tail is observed
if
the hubs are connected by random assortative connections.  The
distribution of travel times for the gradient case shows log-normal behavior
as in the case of the distribution of latencies for the internet
\cite{Sole} and  for
 directed traffic flow \cite{gautam}.

Congestion effects are observed when many messages run simultaneously       
on the base network. However, the network decongests very rapidly when      
the gradient mechanism is applied to a few hubs of high       
co-efficient of betweenness centrality,                        
The existence of transport traps                                     
can set a limit to the extent to which congestion is
cleared at low hub density.
The spatial configuration of traps is studied for both the gradient and
other assortative decongestion schemes.                
We observe that the                                   
gradient mechanism                        
which results in the formation         
of star configurations,                                        
is  substantially less prone to the formation of transport traps                    
than other decongestion mechanisms.                   
We also propose strategies which eliminate the     
trapping effects  either by rerouting, or by minimal addition of capacity     
or connections                                                
at very few locations. These strategies clear the network very
efficiently.  
We note that networks which incorporate geographic
clustering and encounter congestion problems arise in many practical
situations e.g. cellular networks\cite{Jeong}  
and air traffic networks \cite{Sinai}.
Networks where functional clusters are connected by long range
connections arise in  complex brain
networks\cite{kurths} and neural networks\cite{buzsaki} as well.  
Our results may have relevance in these contexts. 
\begin{acknowledgments}
We wish to acknowledge the support of DST, India under the project SP/S2/HEP/10/2003.
\end{acknowledgments}

\end{document}